\documentclass[%
a4paper,
superscriptaddress,
notitlepage,
 amsmath,amssymb,
 aps,
 twocolumn,
 prl
 floatfix
]{revtex4-1}

\usepackage{tikz}
\usetikzlibrary{calc,arrows,decorations.pathreplacing,backgrounds}
\colorlet{mylinkcolor}{blue!66!black!80}
\usepackage{mathtools,amssymb,amsfonts}
\usepackage[colorlinks=true,linkcolor=mylinkcolor,citecolor=mylinkcolor,filecolor=cyan,urlcolor=mylinkcolor,breaklinks=true]{hyperref}
\usepackage{xcolor}

\usepackage[utf8]{inputenc}
\usepackage{booktabs}
\usepackage{bbm,mathrsfs,mathtools}
\usepackage{graphicx}

\newcommand{\e}[1]{\mathrm{e}^{#1}}

\DeclareMathOperator{\sech}{sech}
\DeclareMathOperator{\arctanh}{arctanh}

\DeclareSymbolFontAlphabet{\mathscrsfs}{rsfs}

\newcommand{\blue}[1]{\textcolor{black}{#1}}

\begin{document}
\title{Global Speed Limit for Finite-Time Dynamical Phase Transition and Nonequilibrium Relaxation}

\author{Kristian Blom}
\affiliation{Mathematical bioPhysics group, Max Planck Institute for Multidisciplinary Sciences, G\"{o}ttingen 37077, Germany}

\author{Alja\v{z} Godec}
\email{agodec@mpinat.mpg.de}
\affiliation{Mathematical bioPhysics group, Max Planck Institute for Multidisciplinary Sciences, G\"{o}ttingen 37077, Germany}

\date{\today}

\begin{abstract}
  \noindent Recent works unraveled an intriguing finite-time dynamical
phase transition in the thermal relaxation of the mean field 
Curie-Weiss 
model. The phase transition reflects a sudden switch in the dynamics.
Its existence in systems with a finite range of interaction, however, remained
unclear. \blue{Here we demonstrate the dynamical phase transition
for nearest-neighbor Ising systems on the square and Bethe lattices
through extensive computer 
simulations and by analytical results. Combining
large-deviation techniques and
Bethe-Guggenheim theory 
we prove
the existence of
the dynamical
phase transition for arbitrary
quenches, including those within
the two-phase region.} Strikingly, for any given initial condition
we prove and explain the existence of
non-trivial 
\emph{speed limits} for 
the dynamical
phase transition and the relaxation of magnetization,
which \blue{are fully corroborated by simulations of the microscopic 
  Ising model but} are absent in the mean field setting.~Pair correlations,
which are neglected in mean field theory and trivial in
the Curie-Weiss model,
account for kinetic constraints due to frustrated local configurations that
give rise to a 
global speed limit.
%
\end{abstract}
\maketitle

Despite its overwhelming importance 
in condensed
matter physics \cite{Relaxation,Relaxation_2}, our understanding of
thermal relaxation kinetics is far from complete and mostly limited to systems
near equilibrium \cite{onsager_1,onsager_2,Kubo_1957} and
non-equilibrium  \cite{Seifert_2010,Baiesi_2013,FDT_2020} steady states.\ Notable 
advances
in understanding relaxation dynamics 
out of equilibrium include
far-from-equilibrium fluctuation-dissipation theorems
\cite{Dean,Lippiello}, 
``frenesy'' \cite{Frenesy}, anomalous relaxation a.k.a.\ the Mpemba
effect \cite{raz_mpemba_2017,klich_mpemba_2019,Mpemba_3,Mpemba_4,Oren_2022}, 
optimal heating and cooling \cite{optimal} as well as driving
\cite{ESE,ESE_2} protocols, asymmetries in 
heating and cooling rates \cite{asym,asym_2,asym_3,asym_4},
and dynamical phase transitions (i.e.\ the occurence of non-analytic
points in distributions 
of physical observables)
\cite{Graham_1,Graham_2,Gaw,MFT,Bertini_2010,Cusp_1,Cusp_2,Cusp_3,Garrahan,Garrahan_2,Garrahan_3,Garrahan_4,Ates,Hickey,Peter,Carlo,Hurtado,Hugo,Hurtado_2,Mehl,Speck,Minimal,Peter_2,Rosemary,Peter_3}. Further important results on non-equilibrium relaxation are embodied
in 
thermodynamic uncertainty relations for non-stationary systems
\cite{Pietzonka2017PRE,Dechant2018JPAMT,Liu2020PRL,Koyuk2019PRL,Koyuk2020PRL,Dieball},
and 
so called \emph{speed limits}
\cite{Mandelstam,1983,Aharonov,Pfeifer,Margolus,Nature,Vittorio,Pfeifer,Lutz,Taddei,Campo,Lutz_2,QSL_22,CSL,CSL_2,CSL_3,CSL_4,CSL_5,CSL_6,CSL_7,Massi,CSL_5,Saito,Ito_det}.

In contrast to the well established concept of quantum speed limits
\cite{Mandelstam,1983,Aharonov,Pfeifer,Margolus,Nature,Vittorio,Pfeifer,Lutz,Taddei,Campo,Lutz_2,QSL_22} that has long
been known \cite{Mandelstam}, it was comparably only recently found that
the evolution of classical systems is also bounded by fundamental
speed limits \cite{CSL,CSL_2,CSL_3,CSL_4,CSL_5,CSL_6,CSL_7}. Quantum
and classical speed-limits impose an upper bound on the rate of
change of a system state evolving from a given non-stationary initial state, and arise
as an intrinsic dynamical property of Hilbert space
\cite{CSL}. Moreover, it was found that by considering the thermodynamic cost
of the state change one may derive even sharper thermodynamic speed
limits that bound the rate of
change of a system state from above by the entropy production
rate \cite{CSL_3,Massi,CSL_5,Saito,Ito_det}.

Recently, a 
surprising \emph{finite-time dynamical phase
transition} was observed in a mean field (MF)
Ising system \cite{PhysRevLett.128.110603,meibohm2022landau},
manifested as a finite-time
singularity \cite{singu_1,singu_2} in the probability density of
magnetization \cite{PhysRevLett.128.110603} and entropy flow per spin \cite{meibohm2022landau} upon a
quench from any sub-critical temperature $T<T_c$ to a temperature
$T_q$ \footnote{In \cite{meibohm2022landau} only super-critical quench
temperatures $T_{q}>T_{c}$ are considered, whereas \cite{singu_1,
  singu_2} consider all possible $T_q$.}.~In 
contrast to conventional phase
transitions, 
here time plays
the role of a control parameter inducing an abrupt change of the typical dynamics
\cite{PhysRevLett.128.110603,meibohm2022landau}.~The sudden transition from a
Gibbsian to a non-Gibbsian 
probability density occurs
for \emph{all} quenches 
from sub-critical temperatures $T<T_{c}$, 
whereby the initial location of the singularity depends on $T$ and $T_q$ \cite{singu_2}.~Upon quenches from super-
critical temperatures $T>T_{c}$ the
probability density remains Gibbsian forever \cite{singu_2}, but the dynamics is
non-ergodic \cite{Bray}.

Notwithstanding the detailed results on the non-Gibbsian transition in
the MF setting, it remains unknown if and in what form this novel
dynamical phase transition exists in systems with a finite range of
interactions.  
Moreover, since speed limits
bound from below the time of reaching a final state from a
given initial state, the following intriguing questions arise: 
\emph{What happens with the speed limit in the finite-time dynamical
phase transition, where the dynamics experiences an abrupt change?
Is there a global speed limit to reaching the critical
time?}

To shed light on these questions we here present analytical results on
non-equilibrium relaxation of nearest-neighbor Ising systems on
the Bethe-Guggenheim (BG) level
\cite{bethe1935statistical,guggenheim1935statistical}, which accounts
for nearest-neighbor pair correlations and is exact for the
nearest-neighbor Ising model on the Bethe lattice. \blue{Furthermore,
  we present circumstantial simulation
  evidence for the 
  dynamical phase transition on the
  square and Bethe lattices.}~Our results
confirm, for the first time, 
the existence of the finite-time dynamical phase transition in
finite-range Ising systems. Strikingly, we derive explicit \emph{global} speed limits to
both, the critical time and relaxation time, \blue{on the
  Bethe and square lattices}, which are \blue{fully
  corroborated by simulations of the full Ising model but are} absent in the
MF setting. Notably, the speed limit is set by an antiferromagnetic
interaction and is faster than the dynamics of a non-interacting
system. 
Accounting for kinetically unfavorable local spin configurations, pair correlations, which are neglected in MF
theory, impose a 
global speed limit on the non-Gibbsian dynamical phase transition.


\textit{Fundamentals.---}The Hamiltonian of nearest-neighbor
interacting Ising spins $\sigma_{i}=\pm1, \ i=\{1,...,N\}$ reads
\begin{equation}
    H(\boldsymbol{\sigma},J)=-J\sum_{\langle ij \rangle}\sigma_{i}\sigma_{j},
\end{equation}
with $J$ denoting the ferromagnetic ($J>0$) or antiferromagnetic
($J<0$) coupling and $\langle ij\rangle$ the sum over nearest neighbor
spin pairs.\ The spins are placed on a Bethe lattice with coordination
number $\bar{z}\in\mathbb{N}^{+}$. Let $m(\boldsymbol{\sigma})\equiv
N^{-1}\sum_{i=1}^{N}\sigma_{i}$ be the magnetization per spin for a
given configuration
$\boldsymbol{\sigma}=(\sigma_{1},...,\sigma_{N})$. The equilibrium free energy density in the thermodynamic limit is defined as ${\rm f}(m;J)=-\lim^{m={\rm const.}}_{N\rightarrow\infty}\left[N^{-1}\ln{(\mathcal{Z}_{m}(J))}\right]$,
where
$\mathcal{Z}_{k}(J)\equiv\sum_{\boldsymbol{\sigma}}\exp{(-H(\boldsymbol{\sigma},J)/k_{\rm
    B}T)}\mathbbm{1}_{m(\boldsymbol{\sigma}),k}$ is the
fixed-magnetization partition function
with indicator function
$\mathbbm{1}_{a,b}$ being 1 when $a=b$ and $0$ otherwise. Within
BG theory, 
the free energy density
in units of $k_{\rm
  B}T$, $\tilde{\rm f}_{\rm BG}\equiv{\rm f}_{\rm BG}/k_{\rm
  B}T$, reads (exactly for Bethe lattices) \cite{blom2021criticality,bethe1935statistical,guggenheim1935statistical}
\begin{eqnarray}
    \tilde{\rm f}_{\rm BG}(m;\tilde{J})&=&\bar{z}\tilde{J}(\zeta(m;\tilde{J})-1/2){+}(1{-}\bar{z})[\Xi(m)+\Xi(-m)]\nonumber\\
    &+&\frac{\bar{z}}{2}\sum_{\eta=\pm}[\Xi(\eta m{-}\zeta(m;\tilde{J}))+\Xi(\zeta(m;\tilde{J}){-}1)],
    \label{fBG}
\end{eqnarray}
where $\Xi(m)\equiv(1/2{+}m/2)\ln{(1/2{+}m/2)}$, ${\tilde{J}{\equiv}J/k_{\rm B}}T$, and
\begin{equation}
    \zeta(m;\tilde{J})\equiv\frac{1-m^{2}}{1+[m^{2}+\exp(4\tilde{J})(1-m^{2})]^{1/2}}.
    \label{zetam}
\end{equation}
The MF 
counterpart is recovered by applying the transformation
$\zeta(m;\tilde{J})\rightarrow (1-m^{2})/2$, or equivalently to setting
$\tilde{J}=0$ in Eq.~\eqref{zetam} \footnote{In
\cite{PhysRevLett.128.110603,singu_1,singu_2} there is no explicit
dependence on the lattice coordination number $\bar{z}$. This is
equivalent to setting $\bar{z}=1$ in this work.}.  
The BG critical temperature below which
$\tilde{\rm f}_{\rm BG}(m;\tilde{J})$ develops two degenerate minima
reads $\tilde{J}^{\rm
  BG}_{c}\equiv\ln{(\bar{z}/(\bar{z}-2))}/2$, and correctly diverges
in dimension one with $\bar{z}=2$, where no phase transition
occurs. \blue{An exact result for ${\rm f}(m;\tilde{J})$ on the square lattice
  remains elusive \cite{10.1143/PTP.127.791}, while the critical temperature reads $\tilde{J}^{\rm
  SQ}_{c}\equiv\ln{(1+
  \sqrt{2})}/2$ \cite{PhysRev.65.117}.}

\blue{We focus on the 
magnetization $m$ evolving under Glauber
dynamics of $\boldsymbol{\sigma}$ \cite{glauber_timedependent_1963,blom2021criticality} upon an instantaneous temperature quench ${\tilde{J}_{0}\rightarrow\tilde{J}_{q}<\tilde{J}_{0}}$, where
$\tilde{J}_q$ may be positive or negative.~Let $P_{N}(m;\tilde{J},t)$ be the probability density of $m$ at time
$t$ and ${V_{N}(m;\tilde{J},t)\equiv -N^{-1}\ln
P_{N}(m;\tilde{J},t)}$ the time-dependent large-deviation rate
function.~We set $V(m;\tilde{J},t)\equiv
\lim^{m={
\rm const.}}_{N\rightarrow\infty}V_{N}(m;\tilde{J},t)$.~At equilibrium we have
$V_{\rm
  eq}(m;\tilde{J})\equiv \lim_{t\to\infty}V(m;\tilde{J},t)=\tilde{\rm f}(m;\tilde{J})-\tilde{\rm
  f}(\bar{m};\tilde{J})$ with
$\bar{m}(\tilde{J})\equiv\arg\min_{m}\tilde{\rm f}(m;\tilde{J})$
denoting the location of
free-energy minima. We are interested in the temporal evolution of $V(m;\tilde{J},t)$ upon applying the temperature quench.~Experimentally quenches to
negative $\tilde{J}_q$ may be achieved, e.g.\ by ultrafast optical
switching ferro-antiferromagnetic materials \cite{switching} or by
spin-population inversion in metals by radio-frequency irradiation \cite{Science} yielding 
negative spin temperatures \footnote{Note that only the nuclear spin
temperature becomes negative, other degrees of freedom actually
heat up.~For an excellent pedagogical expose on negative
  temperatures in systems with bounded energy spectra see
  \cite{Frenkel}.}.~Note that quenches beyond the N\'eel point
(i.e.\ the antiferromagnetic critical point) push the
  system across the antiferromagnetic transition, which $m$ does not detect \cite{Ziman_1951,Makoto,Ono,Peruggi_1983}.~In fact, applying the reverse quench and replacing $m$ with the
staggered magnetization \cite{Ziman_1951,Makoto,Ono,Peruggi_1983} yields
    mirror-symmetric results (see \cite{Note4}).}
\begin{figure}[t!]
    \centering
    \includegraphics[width=0.49\textwidth]{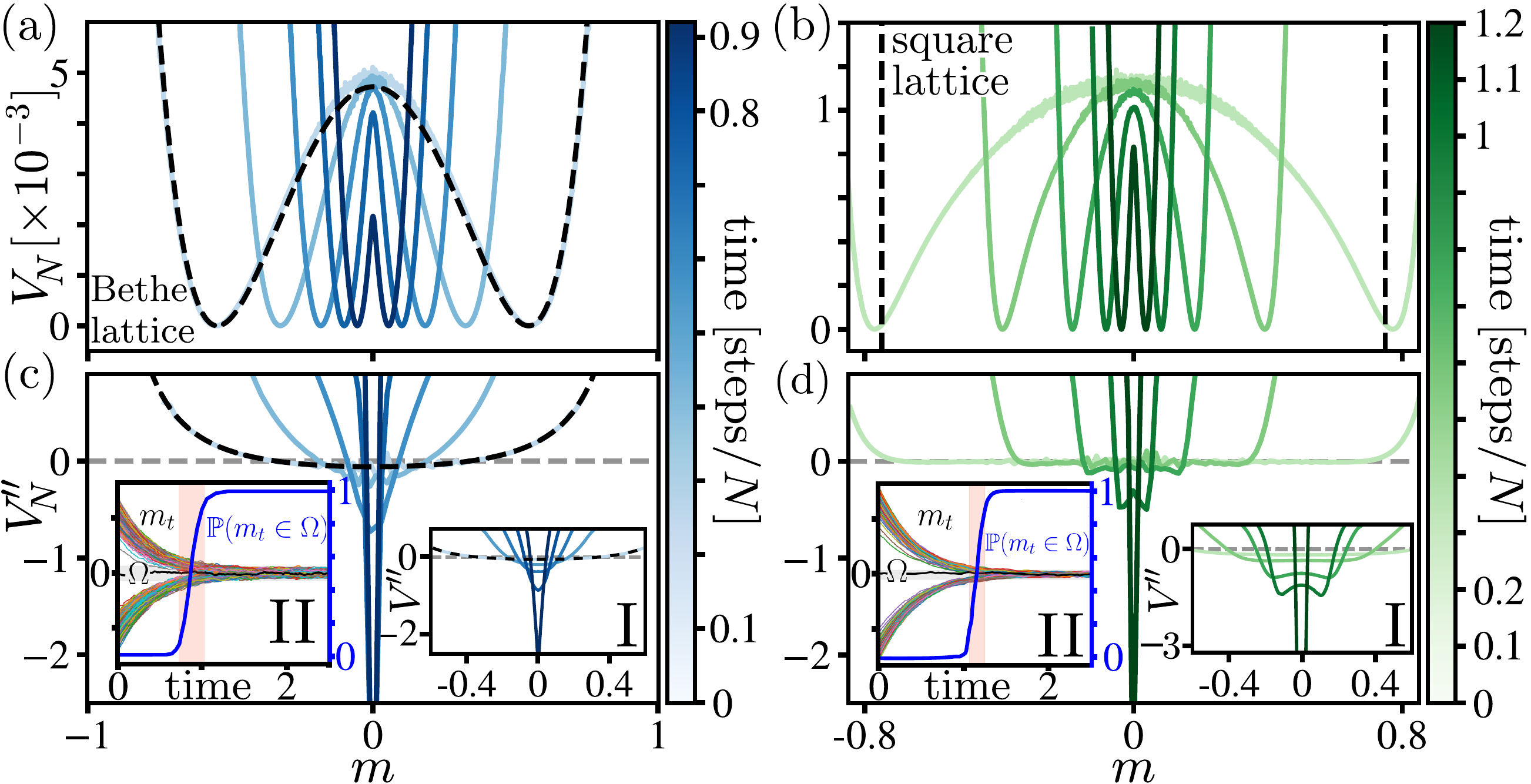}
    \caption{\blue{Kinetic Monte-Carlo (MC) simulations of the
        temporal evolution of $V_{N}(m;\tilde{J},t)$ (a-b) and
        $V_{N}^{\prime\prime}(m;\tilde{J},t)$ (c-d) for a Bethe
        ($\bar{z}=5, N=2000$) (a,c) and square lattice ($N{=}90\times90$) (b,d) upon a
        quench from $\tilde{J}_0=0.275$ (Bethe) and $\tilde{J}_{0}=0.45$ (square) to an
        antiferromagnetic $\tilde{J}_{q}=-0.2$ (see \cite{Note4} for
        simulation details). Time is expressed as the number of MC
        steps per spin, and increases from bright to dark. Black
        dashed lines show the initial equilibrium 
        $V(m;\tilde{J}_{q},0)$ in Eq.~\eqref{fBG} (a) and the minima
        of $V(m;\tilde{J}_{q},0)$ given by Onsager's spontaneous
        magnetization \cite{PhysRev.85.808} (b).~Inset I:~Theoretical
        evolution of $V^{\prime\prime}(m;\tilde{J},t)$ for a lattice
        with $\bar{z}=5$ (c) and $\bar{z}=4$ (d);~profiles are shown 
at equal times as for simulations, but with ${\rm MC \ steps}/N$ replaced by $t/\tau_{r}$, where $\tau_{r}$ is the relaxation time (see text).~Inset II:~Trajectories of the magnetization $m_{t}$ (coloured lines) and the occupation probability $\mathbbm{P}(m_{t}\in\Omega)$  (blue lines) with $\Omega\equiv[-\epsilon,\epsilon]$ with $\epsilon = 0.065$ (left) and $\epsilon=0.04$ (right).}}
    \label{fig1}
    \vspace{-0.35cm}
\end{figure}

\blue{\textit{Simulations.---} We performed discrete-time single
  spin-flip Glauber
  \cite{glauber_timedependent_1963,blom2021criticality} Monte Carlo
  (MC) simulations of the Ising model on the Bethe ($N=2000,
  \ \bar{z}=5$) and square lattice ($N=90\times90$),~MF results are shown in \cite{Note4}. Simulations on the Bethe lattice were performed with the random graph algorithm \cite{D_A_Johnston_1998, Deepak_Dhar_1997}. 
Starting from a random configuration with $m=0$ we equilibrated the
system at temperature $\tilde{J}_{0}>\tilde{J}_{c}$. Upon complete
equilibration (see \cite{Note4} for benchmarks) we changed the
temperature to $\tilde{J}_{q}$ and let the system relax. The
magnetization was sampled at different time points and histograms and
corresponding rate functions (see Fig.~\ref{fig1}a-b) were determined
from an ensemble of $4\times 10^{8}$ (Bethe) and $10^{8}$ (square) independent trajectories.}

\blue{A clear signature that initial equilibration was complete is
  the agreement of the initial rate function with Eq.~\eqref{fBG}
  for the Bethe lattice (Fig.~\ref{fig1}a; black dashed line).~Note
  that the small offset of the barrier diminishes for increasing
  system size (see \cite{Note4}).~Similarly, the minima of the initial
  rate function for the square lattice match Onsager's spontaneous
  magnetization \cite{PhysRev.85.808} (Fig.~\ref{fig1}b; vertical
  black dashed line), where the agreement steadily improves for growing system sizes (see \cite{Note4}).}

\blue{Following $V_{N}$ in time we observe 
  the occurrence of a peak at $m=0$ (see Fig.~\ref{fig1}a-b). To
  evaluate this systematically, we determine the local curvature
  $V_{N}^{\prime\prime}$ (see Fig.~\ref{fig1}c-d and \cite{Note4} for
  details).~Indeed, at some point the curvature near $m=0$ rapidly
  drops to large negative values, thus providing the first
  circumstantial
  evidence for the finite-time dynamical phase transition in 
  nearest-neighbor Ising systems. The abrupt appearance of a
  \emph{true} singularity is, of course, precluded by the finite system size. The simulated curvature
  profiles agree qualitatively with theoretical predictions in the
  $N\to\infty$ limit where the singularity indeed emerges (see Fig.~\ref{fig1}c-d inset I).}

\textit{Theory.---} \blue{To go beyond finite-system MC simulations,} we determined the temporal evolution of $V(m;\tilde{J},t)$ within the \emph{local equilibrium} approximation
\cite{PhysRev.145.224,Kadanoff}, which is highly accurate in the
thermodynamic limit \cite{blom2021criticality}. 
Let $W^{\pm}(M;\tilde{J})$ denote the transition rate to change the
total magnetization from $M\equiv N m\rightarrow M\pm2$ by a
single-spin flip. Following \cite{PhysRev.145.224,Kadanoff} we
define, in the thermodynamic limit, an intensive transition rate
$w^{\pm}(m;\tilde{J})\equiv\lim^{m={\rm
    const.}}_{N\rightarrow\infty}[W^{\pm}(Nm;\tilde{J})/N]$, which reads
\begin{equation}
    w_{{\rm BG}}^{\pm}(m;\tilde{J})=\frac{1\mp m}{2\tau}\left(\e{-\tilde{J}}+\frac{2\zeta(m;\tilde{J})\sinh{(\tilde{J}})}{1\mp m}\right)^{\bar{z}},
    \label{wBG}
\end{equation}
$\tau$ being an intrinsic time-scale of infinitesimal changes of
magnetization $m\to m+\mathrm{d}m$ \cite{Saito1976}. 
The transition rates obey the parity symmetry $w^{\pm}_{\rm
  BG}(m;\tilde{J})=w^{\mp}_{\rm BG}(-m;\tilde{J})$ and detailed
balance w.r.t.~the free energy density, $w^{+}_{\rm
  BG}(m;\tilde{J})/w^{-}_{\rm
  BG}(m;\tilde{J})=\exp{(-2\partial_{m}\tilde{\rm f}_{\rm
    BG}(m;\tilde{J}))}$. In the weak coupling (or high
temperature) limit we recover MF rates
$\lim_{\tilde{J}\rightarrow0}w^{\pm}_{\rm
  BG}(m;\tilde{J})=w^{\pm}_{\rm
  MF}(m;\tilde{J})+\mathcal{O}(\tilde{J}^{2})$ reported in
\cite{PhysRevLett.128.110603}. Out of equilibrium the rate function
$V(m;\tilde{J},t)$
obeys a Hamilton-Jacobi equation \cite{PhysRevLett.128.110603, doi:10.1063/1.467139, PhysRevE.72.046114}
\begin{equation}
    \partial_{t}V(m;\tilde{J},t)+\mathcal{H}(m,\partial_{m}V(m;\tilde{J},t))=0,
    \label{PDE}
\end{equation}
with the Hamiltonian given by 
\begin{equation}
    \mathcal{H}(q,p)=w^{+}(q;\tilde{J})(\e{2p}-1)+w^{-}(q;\tilde{J})(\e{-2p}-1). 
\end{equation}
Eq.~\eqref{PDE} can be derived from the master equation for
$P_{N}$ as the instanton solution in the thermodynamic
limit.
\begin{figure*}[t!]
    \centering
    \includegraphics[width=0.85\textwidth]{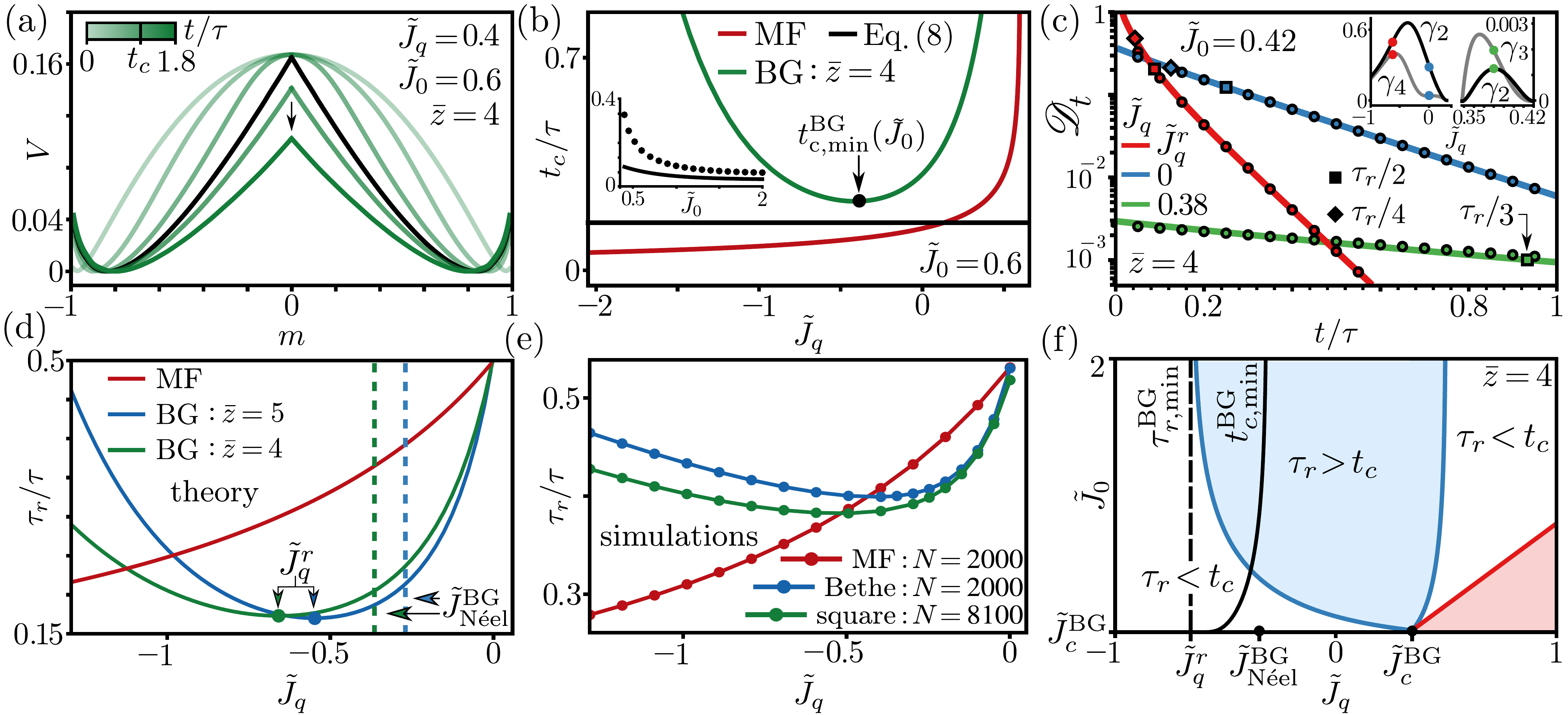}
    \caption{
    (a) Temporal evolution of the BG rate function
      \blue{$V(m;\tilde{J}_{q},t)$} upon a quench into the two-phase
      domain.\blue{~(b) Critical
      time $t_{c}/\tau$ as a function of 
      $\tilde{J}_{q}$ for
        BG (green) and MF (red) theory.~The BG critical time attains a global minimum $t^{\rm BG}_{c, {\rm min}}$ (black dot) for an antiferromagnetic quench, bounded from below by Eq.~\eqref{tmin} (black line);~Inset: $t^{\rm BG}_{c, {\rm min}}$ (black dots) and
      Eq.~\eqref{tmin} (black line) as a function of $\tilde{J}_{0}$.~(c) Temporal evolution of the relative entropy per spin $\mathscrsfs{D}_{t}$ upon a
      quench into the one-phase (red and blue) and two-phase regime (green).~At $\tilde{J}^{r}_{q}$ (red; Eq.~\eqref{Jqr}) the relative entropy relaxes the fastest.~Dots depict analytical results obtained with the first two nonzero terms in Eq.~\eqref{powser2}.~Lines
      correspond to numerical results.~Squares/diamonds
      denote the first $\tau_{r}/2$ and second $\tau_{r}/4$ -
      $\tau_{r}/3$ relaxation time-scales, respectively.~Inset: First
      two nonzero prefactors $\gamma_{k}$ which enter
      Eq.~\eqref{powser2}.~(d) BG ($\bar{z}=4$ green, $\bar{z}=5$ blue) and MF (red) relaxation time
      $\tau_{r}/\tau$ as a function of 
      $\tilde{J}_{q}$.~The BG relaxation time has a local minimum at
      $\tilde{J}^{r}_{q}<0$ (see Eq.~\eqref{Jqr}).~(e) Kinetic MC
      results of the relaxation time $\tau_{r}/\tau$ for the Bethe
      lattice with $\bar{z}=5$ (blue), square lattice (green), and MF lattice (red); see \cite{Note4} for simulation details.~(f) Dynamical phase diagram for 
      $t^{\rm BG}_{c}$ and $\tau^{\rm BG}_{r}$.~The
      red area is forbidden since $\tilde{J}_{0}>\tilde{J}_{q}$.~Dashed/solid black lines
      denote the fastest relaxation and critical time.}}
    \label{fig2}
\end{figure*}

\indent \textit{Dynamical phase transition.---}We assume that the
system is initially prepared at equilibrium in the two-phase regime
$\tilde{J}_{0}>\tilde{J}^{\rm BG}_{c}$ (i.e.\ below the Curie
temperature), and thus  $V_{\rm BG}(m;\tilde{J},0)=\tilde{\rm f}_{\rm
  BG}(m;\tilde{J}_{0}){-}\tilde{\rm f}_{\rm
  BG}(\bar{m};\tilde{J}_{0})$.\ At $t=0$ we apply an instantaneous
quench 
$\tilde{J}_{q}<\tilde{J}_0$ \blue{whereupon} $V_{\rm BG}(m;\tilde{J},t>0)$ evolves according to Eq.~\eqref{PDE}, which we solve numerically (see
Fig.~\ref{fig2}a for a quench with $\tilde{J}_{q}\geq\tilde{J}^{\rm BG}_{c}$).
As $V_{\rm BG}(m;\tilde{J},t)$ relaxes, there is a defined moment $t^{\rm
  BG}_{c}(\tilde{J}_{0},\tilde{J}_{q})$---the critical time---where
$V_{\rm BG}(m;\tilde{J},t)$ \emph{abruptly} develops a cusp (\blue{see
  $V''_{\rm
  BG}$ in the 
inset of Figs.~\ref{fig1}c-d and} black line in
Fig.~\ref{fig2}a). \blue{Thus, the probability measure of $m$} becomes
non-Gibbsian---a phenomenon 
coined finite-time dynamical phase transition
\cite{PhysRevLett.128.110603, singu_1, singu_2} \blue{that} 
is hereby confirmed
in nearest-neighbor Ising systems.\\
\indent The reflection symmetry around $m{=}0$ and local rates $w_{{\rm
    BG}}^{+}$ and $w_{{\rm
    BG}}^{-}$ that are
strictly increasing and decreasing, respectively, in an interval
around $m=0$, ensure that the forward and backward probability fluxes
remain perfectly balanced in a region around $m=0$ during a transient
period after the quench \blue{(see Fig.~\ref{fig1}c-d inset II)}. As a result,
${P_N(m\approx 0;\tilde{J},t)}$ is transiently ``locked'' in the initial
state (see Fig.~\ref{fig2}a).~``Fronts'' of net flux towards $m=0$ gradually develop on each side and drift towards the center.~At the
dynamical phase transition the fronts collide and the dynamics switches from \emph{confined in the
wells} to \emph{exploring the free energy barrier}, i.e.\ between the
formation of defects in ordered domains to their (partial)
melting.\\
%
\indent The fact that the cusp appears upon quenches within the two-phase regime, $\tilde{J}^{\rm
  BG}_{c}\leq\tilde{J}_{q}<\tilde{J}_{0}$, implies that the dynamical
phase transition does not require a change in geometry from a
double- to a single-well potential. Moreover, we show (see \footnote{See Supplemental Material at [...] for detailed derivations
and auxiliary results.}) that the initial cusp location
undergoes a symmetry-breaking transition below the
temperature $\tilde{J}_{0}>\tilde{J}^{\rm SB}_{0}(\tilde{J}_{q})$
whereupon \blue{its initial location deviates} from $m=0$. For infinite temperature
quenches the symmetry-breaking temperature converges to
$\tilde{J}^{\rm SB}_{0,{\rm BG}}(0){=}\ln{([\bar{z}{+}1]/[\bar{z}{-}2])}/2$,
which in the MF setting simplifies to 
$\lim_{\bar{z}\rightarrow\infty}\tilde{J}^{\rm SB}_{0,{\rm
    BG}}(0){=}3/(2\bar{z}){+}\mathcal{O}(1/\bar{z}^2)$ \cite{singu_1,singu_2}.\\
\indent \textit{Critical time.---}We now determine the critical time $t_{\rm c}$,
i.e.\ the first instance a cusp appears at $m=0$.
The critical time can be determined from the curvature
\cite{PhysRevLett.128.110603} or slope \cite{singu_1, singu_2} at
$m=0$ and reads (see derivation in \cite{Note4}) 
\begin{eqnarray}
    t_{\rm c}(\tilde{J}_{0},\tilde{J}_{q}){=}\frac{\ln{(1{-}\tilde{\rm f}^{\prime\prime}(0;\tilde{J}_{q})/\tilde{\rm f}^{\prime\prime}(0;\tilde{J}_{0}))}}{8w^{\pm}(0;\tilde{J}_{q})\tilde{\rm f}^{\prime\prime}(0;\tilde{J}_{q})},
    \label{tc}
\end{eqnarray}
where $\tilde{\rm f}^{\prime\prime}(0;\tilde{J})\equiv {\rm
  d}^{2}\tilde{\rm f}(m;\tilde{J})/{\rm d}m^{2}|_{m=0}$ and all
appearing quantities are given in Eqs.~\eqref{fBG}-\eqref{wBG}.
\blue{Inserting} the MF
free energy density and transition rates in Eq.~\eqref{tc} we recover
the results derived in \cite{PhysRevLett.128.110603,singu_1,singu_2}.
The BG (\blue{green}) and MF (red) critical times as a function of
$\tilde{J}_{q}$ are shown in Fig.~\ref{fig2}b for
$(\bar{z},\tilde{J}_{0})=(4,0.6)$ and display starkly dissimilar
behavior. In particular, the BG critical time displays a global
minimum---a \emph{global speed limit}---that is absent in the MF
setting. This implies a dominant role of local spin configurations,
which are accounted for 
in the BG
theory but ignored in MF theory.\\
\indent \textit{Antiferromagnetic quenches bound the critical time.---}The
stationary points of Eq.~\eqref{tc} cannot be determined
analytically. To confirm that the speed limit indeed exists 
we instead derive a lower bound on Eq.~\eqref{tc}. The critical time  is monotonically increasing with $\tilde{J}_{q}$
for $\tilde{J}^{\rm
  BG}_{c}\leq\tilde{J}_{q}<\tilde{J}_{0}$ (see proof in
\cite{Note4}). Thus, for quenches within
the two-phase regime the critical time is bounded from below by $t^{\rm BG}_{c}(\tilde{J}_{0},\tilde{J}^{\rm BG}_{c})=-(\bar{z}-1)^{\bar{z}}/(4(\bar{z}^2-2\bar{z})^{\bar{z}/2}\tilde{\rm f}_{\rm BG}^{\prime\prime}(0;\tilde{J}_{0}))$.\\
\indent For quenches beyond the critical point, i.e.\ $\tilde{J}_{q}<\tilde{J}^{\rm
  BG}_{c}$, we have $-\tilde{\rm f}_{\rm
  BG}^{\prime\prime}(0;\tilde{J}_{q})/\tilde{\rm f}_{\rm
  BG}^{\prime\prime}(0;\tilde{J}_{0})>0$ and we can apply the
inequality $\ln{(1+x)}>2x/(2+x)$ for $x>0$ \cite{10.2307/3615890} to
the numerator of Eq.~\eqref{tc}. Minimizing the result with respect to
$\tilde{J}_{q}$  then yields a speed limit on the critical time
\begin{equation}
  t^{\rm BG}_{c,{\rm min}}(\tilde{J}_{0}){>}
  \frac{\cosh^{\bar{z}}{(\ln{[\e{2\tilde{J}^{\rm BG}_{c}}(\e{{-}2\tilde{J}_{0}}{+}2/\bar{z}{+}\Delta_{\bar{z}}(\tilde{J}_{0}))]}/2)}}{\bar{z}{-}\e{2\tilde{J}^{\rm BG}_{c}}[(\bar{z}{-}4)\e{{-}2\tilde{J}_{0}}{-}4/\bar{z}{-}\Delta_{\bar{z}}(\tilde{J}_{0})]},
    \label{tmin}
\end{equation}
where
$\Delta_{\bar{z}}(\tilde{J}_{0}){\equiv}[8{+}\bar{z}^{2}\e{-4\tilde{J}_{0}}{+}\bar{z}(\bar{z}{-}4)(1{-}2\e{-2\tilde{J}_{0}})]^{\frac{1}{2}}$. The
bound becomes tighter with increasing $\tilde{J}_{0}$ (see inset
Fig.~\ref{fig2}b) and $\bar{z}$ (see \cite{Note4}), and for
$\tilde{J}_{0}\rightarrow\infty$ attains a minimum value
$1/8$ for $\bar{z}=4$ (see \cite{Note4}). 
Notably, the BG critical time
attains a minimum below the N\'eel point \cite{Makoto,Ono} for an antiferromagnetic quench
$\tilde{J}_{q}<\tilde{J}^{\rm
  BG}_{\text{N\'eel}}=-\tilde{J}^{\rm
  BG}_{\rm c}<0$ (see point in Fig.~\ref{fig2}b).~\blue{Simulations display a similar non-monotonic trend for the instance at
  which $V''_N$ attains a minimum 
  (Fig.~S2 in \cite{Note4}).}\\

\indent \textit{Asymptotic measure equivalence.---}Despite the presence of
a cusp in the rate function for all $t>t_{c}$ we now show that $P_{N\to\infty}(m;\tilde{J},t)$ becomes
measure equivalent \cite{PhysRevLett.115.268701,Touchette2015} to the equilibrium Gibbs measure exponentially fast.~We
quantify the distance between the two measures via the instantaneous
excess free energy density $\mathscrsfs{D}_t$ \cite{LEBOWITZ19571,
  RevModPhys.61.981, doi:10.1063/1.4803847, PhysRevE.82.011144,
  PhysRevLett.104.090601, Vaikuntanathan_2009,asym} defined as
the relative entropy 
per
spin 
${\mathscrsfs{D}_t\equiv\lim_{N\rightarrow\infty}N^{-1}D[P_N(m;\tilde{J},t)||P^{\rm
      eq}_N(m;\tilde{J})]}$.~Explicitly,
\begin{eqnarray}
    \mathscrsfs{D}_t&{=}&\lim_{N\rightarrow\infty}{\int_{-1}^{1}}\!\!\!\e{-NV(m;\tilde{J}_{q},t)}[V_{\rm eq}(m;\tilde{J}_{q}){-}V(m;\tilde{J}_{q},t)]dm\nonumber\\
    &\simeq&\sum_{k=2}^{\infty}\gamma_{k}(\tilde{J}_{0},\tilde{J}_{q})\e{-kt/\tau_{r}(\tilde{J}_{q})},
    \label{powser2}
\end{eqnarray}
where the second line was obtained with the saddle-point approximation
(for derivation and explicit prefactors $\gamma_{k}$ see
\cite{Note4}).~\blue{The relaxation rate entering
  Eq.~\eqref{powser2} reads
$1/\tau_{r}(\tilde{J}_{q}){\equiv}4w^{\pm}(\bar{m}_{\infty};\tilde{J}_{q})\tilde{\rm
  f}^{\prime\prime}(\bar{m}_{\infty};\tilde{J}_{q})$ with
$\bar{m}_{\infty}\equiv\arg\min_{m}\tilde{\rm f}(m;\tilde{J}_{q})$.}~The 
evolution of $\mathscrsfs{D}_t$ for various quenches is shown in Fig.~\ref{fig2}c.~Clearly, ${\mathscrsfs{D}_{t\to\infty}\to 0}$, implying
that ${\lim_{t\rightarrow\infty}V(m;\tilde{J}_{q},t)=V_{\rm
  eq}(m;\tilde{J}_{q})}$ almost everywhere, i.e.\ the large-deviation behavior is ergodic
\cite{PhysRevLett.115.268701,Touchette2015}.

\textit{Antiferromagnetic speed limit for relaxation.---} For quenches beyond the critical point $\tilde{J}_q<\tilde{J}^{\rm
  BG}_{\rm c}$ the relaxation rate depends
non-monotonically on $\tilde{J}_q$ (compare red and green lines in
Fig.~\ref{fig2}c), which is explicitly elaborated in Fig.~\ref{fig2}d
\blue{(theory) and Fig.~\ref{fig2}e (simulations; see \cite{Note4}
  for methods)}. \blue{Qualitatively theory and simulations fully agree, and quantitative differences
  are due to the discrepancy between continuous and discrete time, finite-size effects, \emph{and} the local-equilibrium approximation.} Similarly to $t_{c}$ we find a speed limit, i.e.\ $\tau^{\rm
  BG}_{r}(\tilde{J}_{q})$ is minimal at an antiferromagnetic
quench $\tilde{J}^{r}_{q}$ below the N\'eel point
\begin{equation}
    \tilde{J}^{r}_{q}\!\equiv\!\arg\min_{\tilde{J}_{q}}\tau^{\rm
      BG}_{r}(\tilde{J}_{q})\!=\!\frac{1}{2}\ln{\!\left(\!\frac{\bar{z}-2\sqrt{\bar{z}-1}}{\bar{z}-2}\!\right)}\!<\!\tilde{J}^{\rm
    BG}_{\text{N\'eel}}\,.
    \label{Jqr}
\end{equation}
The
antiferromagnetic speed
limit $\tau_{r}(\tilde{J}^{r}_{q})$ is the result of a trade-off between an
antiferromagnetic interaction deterministically biasing $m$ towards
smaller values on the one hand, and growing kinetic constraints on
energetically accessible local configurations on the other hand.    
When $\tilde{J}_q>\tilde{J}^{\rm
  BG}_{\rm c}$, i.e.\ for quenches within the two-phase regime, there
is no speed limit and
$\tau_{r}$ decreases monotonically with $\tilde{J}_{q}$ towards
zero because quenches become vanishingly small,
$\bar{m}_0-\bar{m}_{\infty}\to 0$ \cite{Note4}.\\

\indent \emph{Dynamical phase diagram.---}Due to asymptotic measure
equivalence the dynamical phase transition may not always be easily
observable, in particular if $t_c>\tau_r$.~In
Fig.~\ref{fig2}f we present a dynamical phase diagram in
the $(\tilde{J}_{0},\tilde{J}_{q})$-plane, showing that the critical time
is not always smaller than the relaxation time.~However, (i) there is
an extended regime where $t_c<\tau_r$ (see blue
region in Fig.~\ref{fig2}f) such that the transition should
be observable and (ii) the (exact) minimal
relaxation time is always smaller than the (exact) smallest critical
time and the latter always lies below the N\'eel
point. The MF phase diagram is, however, starkly different (see \cite{Note4}).\\ 
\indent\emph{Conclusion.---}Our results reveal, for the first time, the finite-time
dynamical phase transition in nearest-neighbor interacting Ising
systems.~Moreover, they unravel non-trivial antiferromagnetic speed limits for the
critical time and the relaxation time of the
magnetization.~\blue{Theoretical results are fully corroborated
  by computer simulations.}~Considering instead quenches
from antiferromagnetically ordered states we in turn find 
mirror-symmetric results for the staggered magnetization \cite{Ziman_1951,Makoto,Ono,Peruggi_1983}.~These unforeseen 
speed limits embody an optimal trade-off between antiferromagnetic
interactions biasing the magnetization towards
smaller values, and a decreasing number of energetically accessible
local configurations that impose kinetic constraints.~As it emerges due to
kinetic constraints imposed by frustrated local
configurations, it should not come as a surprise that the speed limit requires accounting for
nearest-neighbor correlations and is therefore not captured by MF
theory.~Notably, speed limits may also be obtained from
``classical'' \cite{CSL,CSL_2,CSL_3,CSL_4,CSL_5,CSL_6,CSL_7} or
thermodynamic \cite{CSL_3,Massi,CSL_5,Saito,Ito_det} speed limits
which, however, is likely to be more difficult as analytical
solutions for probability density functions, in particular at the
critical time, do not seem to be
feasible.~Our findings may provide insight allowing for
optimization of ultrafast optical-switching ferromagnetic materials \cite{switching}.~Finally, our work provokes further intriguing
questions, in particular on the microscopic path-wise understanding of
the dynamical critical time, 
the effect of an external field, 
the existence of heating-cooling
asymmetries \cite{asym,asym_2,asym_3,asym_4} in different regimes and
across phase transitions, and 
optimal driving protocols 
\cite{ESE,ESE_2,Mpemba_4,optimal} that may be relevant for optical-switching
ferromagnets. 

\emph{Acknowledgments.---}We thank Rick Bebon for insightful discussions.~The financial support from the German
Research Foundation (DFG) through the Emmy Noether Program GO 2762/1-2
(to AG) is gratefully acknowledged.
\let\oldaddcontentsline\addcontentsline
\renewcommand{\addcontentsline}[3]{}
\bibliographystyle{apsrev4-2}
\bibliography{Project3.bib}
\let\addcontentsline\oldaddcontentsline
\clearpage
\newpage
\onecolumngrid
\renewcommand{\thesection}{S\arabic{section}}
\renewcommand{\thefigure}{S\arabic{figure}}
\renewcommand{\theequation}{S\arabic{equation}}
\setcounter{equation}{0}
\setcounter{figure}{0}
\setcounter{page}{1}
\setcounter{section}{0}

\begin{center}
\textbf{Supplemental Material for:\\Global Speed Limit for Finite-Time
  Dynamical Phase Transition and Nonequilibrium
  Relaxation}\vspace{0.5cm}
Kristian Blom \& Alja\v{z} Godec\\
\emph{Mathematical bioPhysics group, Max Planck Institute for Multidisciplinary Sciences, Göttingen 37077, Germany}
\end{center}

\noindent In this Supplementary Material (SM) we present details on
\blue{the kinetic Monte-Carlo simulations}, calculations, and mathematical proofs of the claims made in the
Letter. The sections are organized in the order they appear in the Letter.
\tableofcontents
\section{Kinetic Monte-Carlo simulations}\label{MC}
\blue{Recall that the time-dependent large-deviation rate function for
  finite system sizes is given by
  $V_{N}(m;\tilde{J},t)\equiv-N^{-1}\ln{(P_{N}(m;\tilde{J},t))}$,
  where $P_{N}(m;\tilde{J},t)$ is the probability density of $m$ at
  time $t$. Here we provide details on the kinetic Monte-Carlo (MC)
  simulations which we used to determine $V_{N}(m;\tilde{J},t)$
  displayed in Fig.~1 in the Letter, and the relaxation time shown in
  Fig.~2e in the Letter. Furthermore, we show that a \emph{finite-system
  proxi for the critical time} $t_{\rm min}$ -- defined as the
  instance in time where the curvature of the rate function at $m=0$ is minimal -- depends non-monotonically on $\tilde{J}_{q}<0$ on the Bethe and square lattice.}
\subsection{Lattice setup}
\blue{We performed kinetic MC simulations on three different types of
  lattices: (i) the fully-connected mean field (MF) lattice, (ii) the
  Bethe lattice, and (iii) the square lattice. Simulations on the
  Bethe lattice were performed using the random graph algorithm
  \cite{D_A_Johnston_1998SM, Deepak_Dhar_1997SM}, which works as
  follows: Let us consider a Bethe lattice with coordination number
  $\bar{z}$. First, we create a ring of $i=\{1,...,N\}$ spins, where
  each spin $\sigma_{i}$ is connected to spin $\sigma_{i-1}$ and
  $\sigma_{i+1}$.  To create the remaining $\bar{z}-2$ connections we
  randomly pair spins together on the lattice. The final result is a
  random graph with coordination number $\bar{z}$. Note that for each
  trajectory we create a new random graph. For large $N$ it has been
  shown that the Ising model on an ensemble of  random graphs is
  equivalent to the Ising model on a Bethe lattice
  \cite{D_A_Johnston_1998SM}. Indeed, for large $N$ we find perfect
  agreement between the obtained initial rate function
  $V_{N}(m;\tilde{J},0)$ and the Bethe-Guggenheim (BG) free energy
  density as shown in Fig.~1a in the Letter and
  Figs.~\ref{figSM-1}b. For the MF lattice we connect each spin on the
  ring to all other spins, and the resulting rate function is shown in
  Figs.~\ref{figSM-1}a,d.} 
\subsection{Acceptance rate}
\blue{For single spin-flip dynamics let $\{\sigma_{j}\}^{\prime}_{i}$ denote the spin configuration obtained by flipping spin $i$ while keeping the configuration of all other spins fixed,
    i.e.,\ $\{\sigma_{j}\}^{\prime}_{i}\equiv(-\sigma_{i},\{\sigma_{j\neq
      i}\})$. Moreover, let $p_{i}(\{\sigma_{j}\})$ denote the acceptance
    rate from $\{\sigma_{j}\}$ to $\{\sigma_{j}\}^{\prime}_{i}$ and
    $\Delta\mathcal{H}_{i}(\{\sigma_{j}\})\equiv\mathcal{H}(\{\sigma_{j}\}^{\prime}_{i})-\mathcal{H}(\{\sigma_{j}\})$
    the energy difference (in units of $k_{\rm B}T$) associated with the
    transition. Using the Glauber algorithm the acceptance rate for the single spin-flip takes the form \cite{glauber_timedependent_1963SM}
    \begin{equation}
        p_{i}(\{\sigma_{j}\})=1/(1+\e{-\Delta \mathcal{H}_{i}(\{\sigma_{j}\})}).
    \end{equation}}
\subsection{Number of simulated trajectories}
\blue{In the table below we display the number of trajectories which we used to obtain the rate functions shown in Fig.~1 in the Letter and Figs.~\ref{figSM-1}-\ref{figSM0}. Snapshots of the magnetization $m$ were taken during both the equilibrium and quench round at equidistant time points starting at $t=0$ and ending at the final MC step.}
\begin{table}[h!]
        \centering
        \begin{tabular}{|p{1.8cm}||p{1.3cm}|p{2.3cm}|p{2.1cm}|p{2.3cm}|p{1.9cm}|}
            \hline
            \multicolumn{6}{|c|}{simulation numbers} \\
            \hline
            lattice & size ($N$) & equilibration time [MC steps] & quench time [MC steps] & \# trajectories & \# snapshots \\
            \hline 
            mean field & $2000$ & $6\times10^{5}$ & $6000$ & $2\times10^{8}$  & $200$\\
            Bethe & $2000$ & $6\times10^{5}$ & $4000$ & $4\times10^{8}$  & $200$\\
            square & $90\times90$ & $1.9683\times 10^{8}$ & $16200$ & $10^{8}$ & $100$\\
            \hline
        \end{tabular}
        \label{Table I}
\end{table}
\subsection{Equilibration benchmark}
\begin{figure}[t!]
    \centering
    \includegraphics[width=0.95\textwidth]{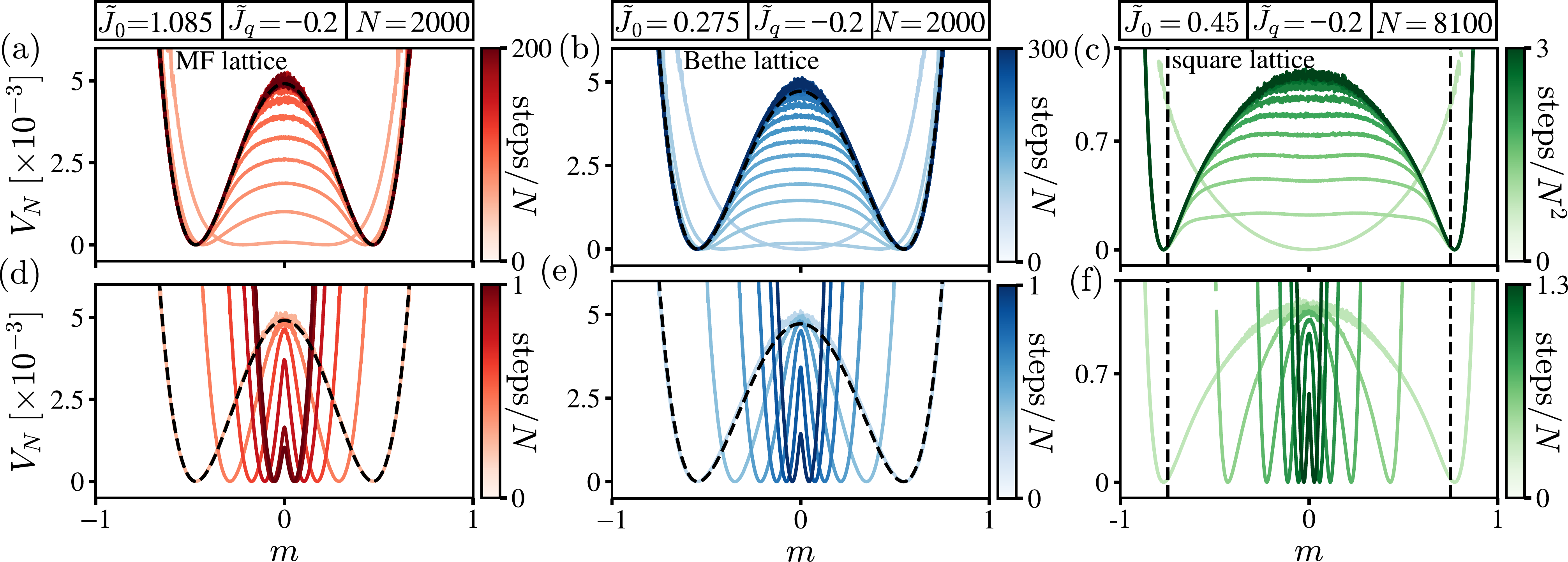}
    \caption{\blue{\textbf{Equilibration and quench dynamics of the rate function.} From left to right we show results for the MF lattice (red), Bethe lattice (blue), and square lattice (green). (a-c) Temporal evolution of $V_{N}(m;\tilde{J},t)$ as a function of $m$ during the equilibration round. (d-f) Temporal evolution of $V_{N}(m;\tilde{J},t)$ as a function of $m$ during the quench round.  Different colors correspond to different times with increasing values from light to dark. Black dashed lines in (a-b, d-e) denote the theoretical equilibrium free energy density for the MF and Bethe lattice, respectively. Vertical black dashed lines in (c, f) correspond to Onsager's spontaneous magnetization given by Eq.~\eqref{mons}.}}
    \label{figSM-1}
\end{figure}

 \blue{Starting from a random configuration with $m=0$ we first
   performed an equilibration round at temperature
   $\tilde{J}_{0}>\tilde{J}_{c}$. To check whether equilibrium was
   reached, we show in Figs.~\ref{figSM-1}a-c the rate function over
   time during the equilibration round. For late times we find that
   the rate functions collapse onto the same curve, which provide a
   first indication that equilibrium is reached. Furthermore, for the
   MF and Bethe lattice we find perfect agreement between the
   numerical rate function for a finite $N$ and the theoretical
   equilibrium result (see black dashed lines in
   Figs.~\ref{figSM-1}a-b). For the square lattice the exact equilibrium free energy density is to date unknown \cite{10.1143/PTP.127.791SM}. However, the locations of the minima are given by Onsager's spontaneous magnetization \cite{PhysRev.65.117SM} (see black dashed vertical lines in Figs.~\ref{figSM-1}c,f) which reads
 \begin{eqnarray}
    \arg\min_{m} V(m;\tilde{J}_{0},0)=\left(1\pm([1-\sinh^{-4}{(2J_{0})}]^{1/8})\right), \ {\rm for} \ J_{0}\geq \frac{1}{2}\ln{(1+\sqrt{2}}).
    \label{mons}
 \end{eqnarray}
 Indeed, in Figs.~\ref{figSM-1}c we find that the minima of the square lattice rate function are located around this value, providing a second indication that equilibrium is reached.}
\subsection{Curvature of the rate function and proxy for the critical time}
\begin{figure}[t!]
    \centering
    \includegraphics[width=0.95\textwidth]{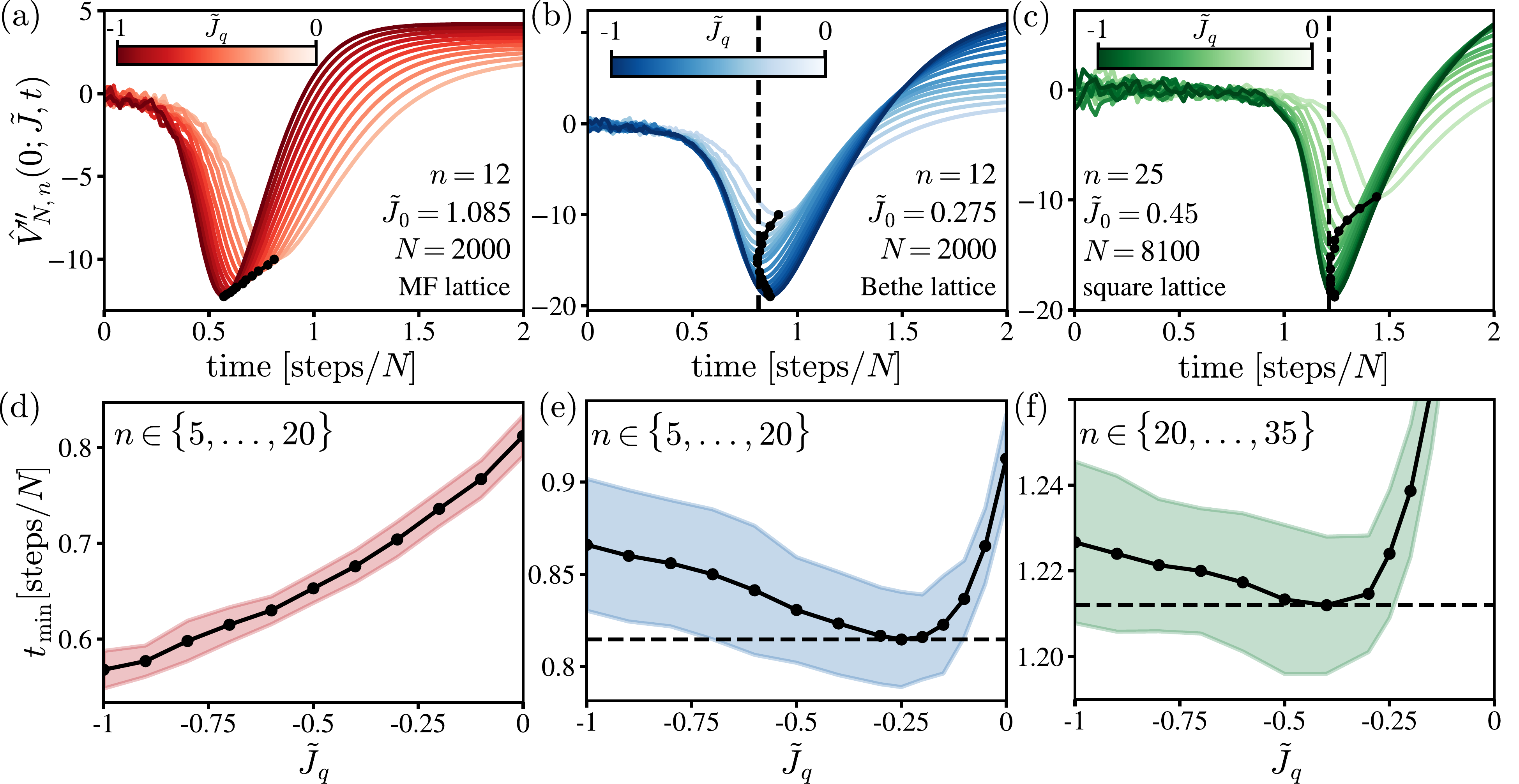}
    \caption{\blue{\textbf{Curvature of the rate function around
          $\mathbf{m=0}$ and proxy for the critical time.} From left to right we
        show results for the MF lattice (red), Bethe lattice (blue),
        and square lattice (green). (a-c) Temporal evolution of
        $V_{N,n}^{\prime\prime}(0;\tilde{J},t)$ (see Eq.~\eqref{Vcg})
        as a function of time $[{\rm steps}/N]$. Different colors
        correspond to different $\tilde{J}_{q}$ with increasing values
        from light to dark. Black dots indicate the minimum of the
        curvature, which we denote by $t_{\rm min}$ and take as a
        proxy for the critical time of the finite-time dynamical phase
        transition. Black dashed lines in (b-c) denote the minimum of $t_{\rm min}$ for the Bethe and square lattice,
        respectively. (d-f) Black line: Averaged critical time over
        various bin sizes $n$ (see range of $n$ in figure) as a
        function of $\tilde{J}_{q}$. Colored shaded area: Standard
        deviation of the critical time over various bin sizes
        $n$. Black dashed lines in (e-f) denote the minimum of $t_{\rm min}$ for the Bethe and square lattice, respectively.}
   }
    \label{figSM0}
\end{figure}

\blue{To evaluate the curvature of the rate function $V_{N}^{\prime\prime}(m;\tilde{J},t)$ used for Fig.~1c,d in the Letter we used the finite difference method. Since the rate function contains strong fluctuations on the level of single spins with resolution $\Delta m=2/N$, we first coarse-grain the rate function through a binning procedure. For a bin size given by $2n+1\in\{1,3,..,\}$ we obtain a coarse-grained rate function in the following way
\begin{equation}
    \hat{V}_{N,n}(m;\tilde{J},t)=-N^{-1}\ln{\left(\frac{1}{2n+1}\sum_{k=-n}^{n}P_{N}(m+2k/N;\tilde{J},t)\right)}, \ {\rm for} \ m\in\{-(N+2n)/N,...,(N-2n)/N\}.
    \label{Vcg}
\end{equation}
Note that the factor $1/(2n+1)$ inside Eq.~\eqref{Vcg} keeps the coarse-grained probability density normalized. After coarse-graining we obtain the curvature with the higher-order finite-difference method
\begin{equation}
    \hat{V}_{N,n}^{\prime\prime}(m;\tilde{J},t)\approx \frac{\sum_{\eta = \pm}[-\hat{V}_{N,n}(m+4\eta/N;\tilde{J},t)+16\hat{V}_{N,n}(m+2\eta/N;\tilde{J},t)-15\hat{V}_{N,n}(m;\tilde{J},t)]}{12(2n/N)^2}+\mathcal{O}((2n/N)^4).
\end{equation} 
In Fig.~\ref{figSM0}a-c we show the curvature around $m=0$ as a
function of time $[{\rm steps}/N]$ for the MF lattice (a), Bethe
lattice (b), and square lattice (c). In each of the lattices we find
that the curvature quickly drops to a large negative value at a finite
time. Interpreting the minima as a proxy for the critical time for
finite systems
attained at $t_{\rm min}$, we show in Fig.~\ref{figSM0}d-f that on the Bethe
and square lattice this proxy is non-monotonic in $\tilde{J}_{q}$ for
antiferromagnetic quenches. Furthermore, for the MF lattice the
critical time decreases steadily with stronger antiferromagnetic
quenches. Comparing these results with Fig.~2b in the Letter we find
strong qualitative agreement between the results for $t_{c}$ obtained
with theory and the proxy $t_{\rm min}$ obtained with kinetic MC simulations. }
\subsection{Evaluation of the relaxation time}
\blue{To obtain the relaxation time $\tau_{r}$ from MC simulations we use the result for the relative entropy given by Eq.~(9) in the Letter, i.e.~
\begin{equation}
    \mathscrsfs{D}_t{=}\lim_{N\rightarrow\infty}{\int_{-1}^{1}}\!\!\!\e{-NV(m;\tilde{J}_{q},t)}[V_{\rm eq}(m;\tilde{J}_{q}){-}V(m;\tilde{J}_{q},t)]dm
    \simeq\sum_{k=2}^{\infty}\gamma_{k}(\tilde{J}_{0},\tilde{J}_{q})\e{-kt/\tau_{r}(\tilde{J}_{q})}.
    \label{powser3}
\end{equation}
Replacing the integral over $m$ in Eq.~\eqref{powser3} by a sum, we compute the relative entropy with the rate functions obtained from the MC simulations. To extract the relaxation time $\tau_{r}$ we make use of the exponential series on the right hand side of Eq.~\eqref{powser3} and take the long-time limit, which gives
\begin{equation}
    \lim_{t\rightarrow\infty}\frac{\ln{(\mathscrsfs{D}_t)}}{t}=
    \lim_{t\rightarrow\infty}\frac{\ln{(\gamma_{2}(\tilde{J}_{0},\tilde{J}_{q})\e{-2t/\tau_{r}(\tilde{J}_{q})})}}{t}=-\frac{2}{\tau_{r}}.
\label{rel_t}    
\end{equation}
Plugging $\mathscrsfs{D}_t$ inferred from simulations into
Eq.~\eqref{rel_t} we extract the relaxation time as shown in Fig.~2e in the Letter.}
\section{Hamiltonian formalism of large deviation function}\label{Ham}
\noindent Recall that $V(m;\tilde{J},t)\equiv
\lim^{m={\rm const.}}_{N\rightarrow\infty}V_{N}(m;\tilde{J},t)$. In the SM
of \cite{PhysRevLett.128.110603SM} it is shown that the rate function
$V(m;\tilde{J}_{q},t)$ with quench temperature
$\tilde{J}_{q}$ obeys the Hamilton-Jacobi (HJ) equation given by
Eq.~(5) in the Letter. The HJ equation can be solved with the \emph{method of characteristics} as follows: Let $\{q(s),p(s)\} \ 0\leq s\leq t$ be the characteristics that solve the Hamilton's equations 
\begin{equation}
        \dot{q}(s)=\partial_{p}\mathcal{H}(q,p), \ \dot{p}(s)=-\partial_{q}\mathcal{H}(q,p), \ q(t)=m, \ p(0)=\tilde{\rm f}^{\prime}(q(0);\tilde{J}_{0}),
        \label{Hamilton}
\end{equation}
where $\dot{q}(s)\equiv dq(s)/ds$, $\dot{p}(s)\equiv dp(s)/ds$, $\tilde{\rm f}^{\prime}(a;\tilde{J})\equiv\partial_{m}\tilde{\rm f}(m;\tilde{J})|_{m=a}$, and $\mathcal{H}(q,p)$ is given by Eq.~(6) in the Letter. Upon solving the Hamilton's equations, the solution to the HJ equation reads
\begin{equation}
    V(m;\tilde{J}_{q},t)=\int_{0}^{t}[p(s)\dot{q}(s)-\mathcal{H}(q,p)]ds+V(q(0);\tilde{J}_{0},0).
    \label{Vsol}
\end{equation}
For $t>t_{c}$, where $t_{c}=t_{c}(\tilde{J}_{0},\tilde{J}_{q})$ denotes the critical time, the solutions to the Hamilton's equations become degenerate. Under these circumstances, the solution that minimizes Eq.~\eqref{Vsol} corresponds to the stable solution \cite{meibohm2022landau2}. 
\section{Lagrangian formalism of large deviation function}\label{Lan}
\noindent One can also obtain the solution to the HJ equation with the Lagrangian formalism, which is formally introduced in \cite{singu_1SM,singu_2SM}. The Lagrangian is obtained from the Hamiltonian via the backward Legendre transform $\mathcal{L}(q,\dot{q})=p(q,\dot{q}) \dot{q}-\mathcal{H}(q,p(q,\dot{q}))$,
where $p(q,\dot{q})$ can be obtained from the first of the Hamilton's equations in Eq.~\eqref{Hamilton} and reads
\begin{equation}
    p(q,\dot{q})=\frac{1}{2}\ln{\left(\frac{\dot{q}+\Lambda(q,\dot{q})}{4w^{+}(q;\tilde{J}_{q})}\right)},
\end{equation}
with $\Lambda(q,\dot{q})\equiv[16w^{+}(q;\tilde{J}_{q})w^{-}(q;\tilde{J}_{q})+\dot{q}^2]^{1/2}$.
Plugging this expression back into $\mathcal{H}(q,p(q,\dot{q}))$ we obtain the Lagrangian
\begin{equation}
    \mathcal{L}(q,\dot{q})=p(q,\dot{q})\dot{q}-\Lambda(q,\dot{q})/2+w^{+}(q;\tilde{J}_{q})+w^{-}(q;\tilde{J}_{q}).
\end{equation}
The Hamilton's equations are replaced by the Euler-Lagrange (EL) equation, which reads
\begin{equation}
    \ddot{q}(s)=2\Lambda(q,\dot{q})\partial_{q}[w^{+}(q;\tilde{J}_{q}){+}w^{-}(q;\tilde{J}_{q})]{-}8\partial_{q}w^{+}(q;\tilde{J}_{q})w^{-}(q;\tilde{J}_{q}), \
    \dot{q}(0)=g(q(0)), \ 
    q(t)=m.
    \label{E-L}
\end{equation}
The boundary condition for $\dot{q}(0)$ is determined by the \emph{curve of allowed initial configurations} (see also Eq.~(24) in \cite{singu_2SM})
\begin{equation}
    g(m)\equiv2\exp{(2\tilde{\rm f}^{\prime}(m;\tilde{J}_{0}))}w^{+}(m;\tilde{J}_{q})-2\exp{(-2\tilde{\rm f}^{\prime}(m;\tilde{J}_{0}))}w^{-}(m;\tilde{J}_{q}),
    \label{g}
\end{equation}
which will be used in Sec.~\ref{symmetry-breaking} to determine the
symmetry-breaking transition. 
Upon solving the EL equation, the solution of the HJ equation is given by 
\begin{equation}
    V(m;\tilde{J}_{q},t)=\int_{0}^{t}\mathcal{L}(q(s),\dot{q}(s))ds+V(q(0);\tilde{J}_{0},0),
    \label{V2}
\end{equation}
which is identical to Eq.~\eqref{Vsol}. Similar to the Hamiltonian
formalism, the solution of Eq.~\eqref{E-L} becomes degenerate for
$t>t_{c}$. The stable solution for $q(s)$ minimizes the rate function
given by Eq.~\eqref{V2}. 
\section{Derivation of the critical time}
\noindent In this section we derive the critical time $t_{c}$ based on
two different approaches which are discussed in
\cite{PhysRevLett.128.110603SM} and \cite{singu_2SM},
respectively. The first approach uses the Hamiltonian formalism
discussed in Sec.~\ref{Ham} to derive an equation for the curvature at
$m=0$.  The second approach uses an invariance principle for the
solutions of Eq.~\eqref{E-L}  discussed in Sec.~\ref{Lan}.   Both
approaches lead to the same result for the critical time given by
Eq.~(7) in the Letter.  However, with the latter approach we can
also derive the initial temperature below which the initial location
of the cusp deviates from $m=0$.
\subsection{Hamiltonian formalism and the Ricatti equation}
\noindent \noindent The critical time
$t_{c}(\tilde{J}_{0},\tilde{J}_{q})$ is defined as the moment when the
rate function $V(m;\tilde{J}_{q},t)$ \emph{develops a cusp at $m=0$},
leading to a negatively diverging curvature. In the SM of
\cite{PhysRevLett.128.110603SM} an equation for the curvature
$V_{0}^{\prime\prime}(\tilde{J}_{q},t)\equiv
V^{\prime\prime}(0;\tilde{J}_{q},t)$ is derived from the Hamilton's
equations.  The resulting equation -- after simplification -- reads
\begin{equation}
    \frac{dV^{\prime\prime}_{0}(\tilde{J}_{q},t)}{dt}=8w^{\pm}(0;\tilde{J}_{q})V^{\prime\prime}_{0}(\tilde{J}_{q},t)(\tilde{\rm f}^{\prime\prime}(0;\tilde{J}_{q})-V_{0}^{\prime\prime}(\tilde{J}_{q},t)),
    \label{Ricatti}
\end{equation}
with initial condition
$V_{0}^{\prime\prime}(\tilde{J}_{q},0)=\tilde{\rm
  f}^{\prime\prime}(0;\tilde{J}_{0})$. To obtain Eq.~\eqref{Ricatti}
we explicitly used the detailed-balance relation
$\ln{(w^{-}(m;\tilde{J})/w^{+}(m;\tilde{J}))}{=}2\tilde{\rm
  f}^{\prime}(m;\tilde{J})$ and the parity symmetry
$w^{\pm}(m;\tilde{J}){=}w^{\mp}(-m;\tilde{J})$ to write
$\partial_{m}w^{\pm}(m;\tilde{J})|_{m=0}=\mp
w^{\pm}(0;\tilde{J})\tilde{\rm f}^{\prime\prime}(0;\tilde{J})$.
Eq.~\eqref{Ricatti} is a so-called \emph{Ricatti equation}, which can
be solved analytically. The resulting solution \emph{up to the critical time} reads
\begin{eqnarray}
    V^{\prime\prime}_{0}(\tilde{J}_{q},t)=\frac{\tilde{\rm f}^{\prime\prime}(0;\tilde{J}_{q})}{1-(1-\tilde{\rm f}^{\prime\prime}(0;\tilde{J}_{q})/\tilde{\rm f}^{\prime\prime}(0;\tilde{J}_{0}))\e{-2t/\hat{\tau}_{r}(\tilde{J}_{q})}},
    \label{V}
\end{eqnarray}
where $1/\hat{\tau}_{r}(\tilde{J}_{q})\equiv
4w^{\pm}(0;\tilde{J}_{q})\tilde{\rm
  f}^{\prime\prime}(0;\tilde{J}_{q})$ is an effective relaxation
rate. The critical time $t_{c}$ determines the root of the denominator
in Eq.~\eqref{V}. Solving for the root leads to Eq.~(7) in the main
Letter.  
\begin{figure}[t!]
    \centering
    \includegraphics[width=\textwidth]{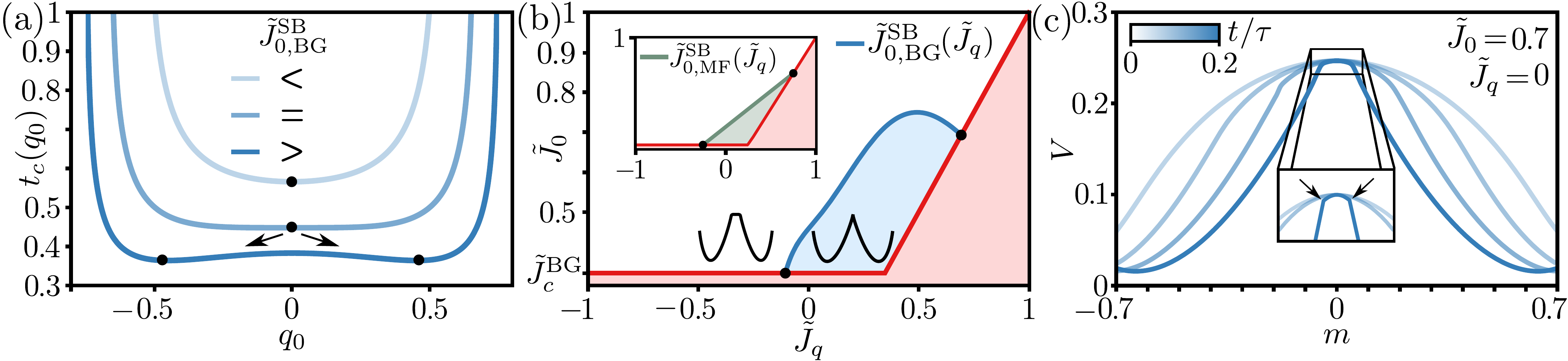}
    \caption{\textbf{Symmetry-breaking transition for the location of
        the cusp.} In all panels we consider a lattice with
      $\bar{z}=4$.  (a) BG critical time $t^{\rm BG}_{c}(q_{0})$ given
      by Eq.~\eqref{tcq0} as a function of the initial point $q_{0}$
      for various values of the initial temperature
      $\tilde{J}_{0}$. The black dots indicate the minima of $t^{\rm
        BG}_{c}(q_{0})$, which set the location of the cusp. For
      $\tilde{J}_{0}>\tilde{J}^{\rm SB}_{\rm 0, BG}$ the critical time
      contains two minima $\pm q_{\rm min}$ (black dots), which
      correspond to non-zero cusp locations. (b) Blue line: BG
      symmetry-breaking temperature $\tilde{J}^{\rm
        SB}_{0}(\tilde{J}_{q})$ given by Eq.~\eqref{JsBG} as a
      function of the quench temperature $\tilde{J}_{q}$. Inside the
      light blue region the cusp is formed at $m=0$, and in the white
      region the cusp is formed at $m\neq0$. The red area is forbidden
      since $\tilde{J}_{0}>\tilde{J}^{\rm BG}_{c}$ and
      $\tilde{J}_{q}<\tilde{J}_{0}$. Inset: MF symmetry-breaking
      temperature $\tilde{J}^{\rm MF}_{0}(\tilde{J}_{q})$ given by
      Eq.~\eqref{JsMF}.  Inside the light green region the cusp is
      formed at $m=0$.  (c) Temporal evolution of the BG rate function
      $V_{\rm BG}(m;\tilde{J}_{q},t)$ for a quench to
      $\tilde{J}_{q}=0$. Time increases from light to dark blue. The
      initial temperature is set below the symmetry-breaking
      temperature, i.e.~$\tilde{J}_{0}>\tilde{J}^{\rm
        SB}_{0}(\tilde{J}_{q})$, to induce a cusp at $m\neq0$. Inset:
      Enlargement of the rate function around the center. Black arrows
      indicate the location of the cusps. } 
    \label{figSM1}
\end{figure}
\subsection{Lagrangian formalism and the symmetry-breaking transition}\label{symmetry-breaking}
\noindent Following the steps in Sec.~3.5 of \cite{singu_2SM} we can
derive the critical temperature $\tilde{J}^{\rm
  SB}_{0}(\tilde{J}_{q})$,  below which the initial location of the
cusp deviates from $\bar{m}=0$. The idea behind this calculation is that at the critical time the solution of Eq.~\eqref{E-L} converges to the same point $q(t_{c})$ for \emph{different initial conditions} $\{q(0),\dot{q}(0)\}$. In other words, \emph{the location of $q(t_{c})$ remains invariant under a variation of the initial conditions}. To determine the symmetry-breaking transition it suffices to consider the dynamics of $q(s)$ around the origin \cite{singu_2SM}. We linearize Eq.~\eqref{E-L} around the point $(q,\dot{q})=(0,0)$, which yields
\begin{eqnarray}
    \ddot{q}(s)&=&q(s)/\hat{\tau}_{r}^2(\tilde{J}_{q}), \ \dot{q}(0)=g(q_{0})\equiv v_{0}, \ q(0)\equiv q_{0},
    \label{E-L2}
\end{eqnarray}
where $\{q_{0},v_{0}(q_{0})\}$ are the initial conditions, and $1/\hat{\tau}_{r}(\tilde{J}_{q})\equiv 4w^{\pm}(0;\tilde{J}_{q})\tilde{\rm f}^{\prime\prime}(0;\tilde{J}_{q})$. The solution of Eq.~\eqref{E-L2} is given by
\begin{equation}
    q(s)=(q_{0}/2-\hat{\tau}_{r}v_{0}/2)\e{-s/\hat{\tau}_{r}}+(q_{0}/2+\hat{\tau}_{r}v_{0}/2)\e{s/\hat{\tau}_{r}}.
\end{equation}
We now consider a variation of $q(s)$ w.r.t.~the initial conditions $\{q_{0},v_{0}(q_{0})\}$, which gives
\begin{equation}
    \frac{dq(s)}{dq_{0}}=\frac{\partial q(s)}{\partial q_{0}}+\frac{\partial q(s)}{\partial v_{0}}g'(q_{0})=(1/2-\hat{\tau}_{r}g'(q_{0})/2)\e{-s/2\hat{\tau}_{r}}+(1/2+\hat{\tau}_{r}g'(q_{0})/2)\e{s/2\hat{\tau}_{r}},
    \label{var}
\end{equation}
where $g'(q_{0})\equiv dg(m)/dm|_{m=q_{0}}$ and $g(m)$ is given by
Eq.~\eqref{g}. At the critical time $s=t_{c}$ the variation
\eqref{var} vanishes, which leads to the critical time in the form
\begin{equation}
    t_{c}(q_{0})=(\hat{\tau}_{r}/2)\ln{\left(\frac{g'(q_{0})-1/\hat{\tau}_{r}}{g'(q_{0})+1/\hat{\tau}_{r}}\right)}.
    \label{tcq0}
\end{equation}
For $\tilde{J}_{c}<\tilde{J}_{0}<\tilde{J}^{\rm
  SB}_{0}(\tilde{J}_{q})$ the critical time given by Eq.~\eqref{tcq0}
has a single minimum at $q_{\rm min}=0$ (see upper line in
Fig.~\ref{figSM1}a). Inserting $q_{0}=0$ and recalling the relation
$\partial_{m}w^{\pm}(m;\tilde{J})|_{m=0}=\mp
w^{\pm}(0;\tilde{J})\tilde{\rm f}^{\prime\prime}(0;\tilde{J})$  we
obtain the critical time given by Eq.~(7) in the Letter.  
\\
\\
\noindent For $\tilde{J}_{0}>\tilde{J}^{\rm SB}_{0}(\tilde{J}_{q})$
Eq.~\eqref{tcq0} develops two minima at $\pm q_{\rm min}\neq0$,
\emph{corresponding to the new cusp locations} (see lower line in
Fig.~\ref{figSM1}a).  
\\
\\
\noindent For $\tilde{J}_{0}=\tilde{J}^{\rm SB}_{0}(\tilde{J}_{q})$
the curvature of Eq.~\eqref{tcq0} at $q_{0}=0$ vanishes (see middle
line in Fig.~\ref{figSM1}a), which results in the following equation
determining $\tilde{J}^{\rm SB}_{0}(\tilde{J}_{q})$
\begin{equation}
    g^{\prime\prime\prime}(0)|_{\tilde{J}^{\rm SB}_{0}(\tilde{J}_{q})}=0,
    \label{symeq}
\end{equation}
where we have used that $g^{\prime\prime}(0)=0$. Solving Eq.~\eqref{symeq} for the MF approximation we obtain the simple result
\begin{equation}
    \tilde{J}^{\rm SB}_{0,{\rm MF}}(\tilde{J}_{q})=\frac{3+\bar{z}\tilde{J}_{q}}{2\bar{z}}.
    \label{JsMF}
\end{equation}
For $\tilde{J}_{q}=0$ we obtain $\tilde{J}^{\rm SB}_{0,{\rm
    MF}}(0)=3/2z$ as mentioned in \cite{singu_1SM,singu_2SM}. 
For the BG approximation the general formula for $\tilde{J}^{\rm SB}_{0,{\rm BG}}(\tilde{J}_{q})$ is rather long and therefore not shown. For $\bar{z}=4$ the result can compactly be written as
\begin{equation}
    \tilde{J}^{\rm SB}_{0,{\rm BG}}(\tilde{J}_{q})|_{z=4}=\ln{(x_{\tilde{J}_{q}})}/2,
    \label{JsBG}
\end{equation}
where $x_{\tilde{J}_{q}}$ is the \emph{real solution} of the following cubic equation
\begin{equation}
    20{-}16(1{+}2\e{-2\tilde{J}_{q}})x_{\tilde{J}_{q}}{+}(8{+}8\e{-2\tilde{J}_{q}}{+}20\e{-4\tilde{J}_{q}})x_{\tilde{J}_{q}}^2{-}(2{-}4\e{-2\tilde{J}_{q}}{+}10\e{-4\tilde{J}_{q}}{-}\e{-8\tilde{J}_{q}}{+}6\e{-10\tilde{J}_{q}}{-}9\e{-12\tilde{J}_{q}}{+}4\e{-14\tilde{J}_{q}})x_{\tilde{J}_{q}}^3{=}0.
\end{equation}
For $\tilde{J}_{q}=0$ we obtain $\tilde{J}^{\rm SB}_{0,{\rm
    BG}}(0)=\ln{(\frac{\bar{z}+1}{\bar{z}-2})}/2$ as mentioned in the
Letter. In Fig.~\ref{figSM1}b we plot Eq.~\eqref{JsBG} as a
function of $\tilde{J}_{0}$ with the dark blue line. Interestingly,
the light blue region for which the cusp appears at $m=0$ is rather
small and of finite area. Correspondingly, in Fig.~\ref{figSM1}c we
provide an example of the rate function $V_{\rm
  BG}(m;\tilde{J}_{q},t)$ for which the cusps appear at a non-zero
locations.  
\section{Bounds on the BG critical time}
\noindent In this section we derive the bounds for the BG critical time $t^{\rm BG}_{c}$. Inserting the BG free energy density and transition rates -- given by Eqs.~(2) and (4) in the Letter -- into Eq.~(7) in the Letter, we obtain 
\begin{equation}
    t^{\rm BG}_{c}(\tilde{J}_{0},\tilde{J}_{q})=\frac{\cosh^{\bar{z}}{(\tilde{J}_{q})}(\tanh{(\tilde{J}_{q})}+1)}{4((\bar{z}-1)\tanh{(\tilde{J}_{q})}-1)}\left[\tilde{J}_{q}+\ln{\left(\frac{(\bar{z}-1)\sinh{(\tilde{J}_{0})}-\cosh{(\tilde{J}_{0})}}{\bar{z}\sinh{(\tilde{J}_{0}-\tilde{J}_{q})}}\right)}\right],
    \label{tcBG}
\end{equation}
where $\tilde{J}_{0}>\tilde{J}^{\rm BG}_{c}\equiv\ln{(\bar{z}/(\bar{z}-2))}/2$ and $\tilde{J}_{q}\leq\tilde{J}_{0}$.
Fig.~2b in the Letter displays the BG critical given by
Eq.~\eqref{tcBG} with the green line. The BG critical time has a
minimum for an anti-ferromagnetic quench $\tilde{J}_{q}<0$, which
cannot be determined analytically. We can, however, derive lower
bounds on the critical time. To construct the bounds we will
distinguish between quenches in the one- and two-phase domain,
i.e. $\tilde{J}_{q}<\tilde{J}^{\rm BG}_{c}$ and
$\tilde{J}_{q}\geq\tilde{J}^{\rm BG}_{c}$. The general result for the
anti-ferromagnetic bound is given by Eq.~(8) in the Letter.\\
\flushleft \subsection{$\tilde{J}_{q}<\tilde{J}^{\rm BG}_{c}$}
\begin{figure}[t!]
    \centering
    \includegraphics[width=0.67\textwidth]{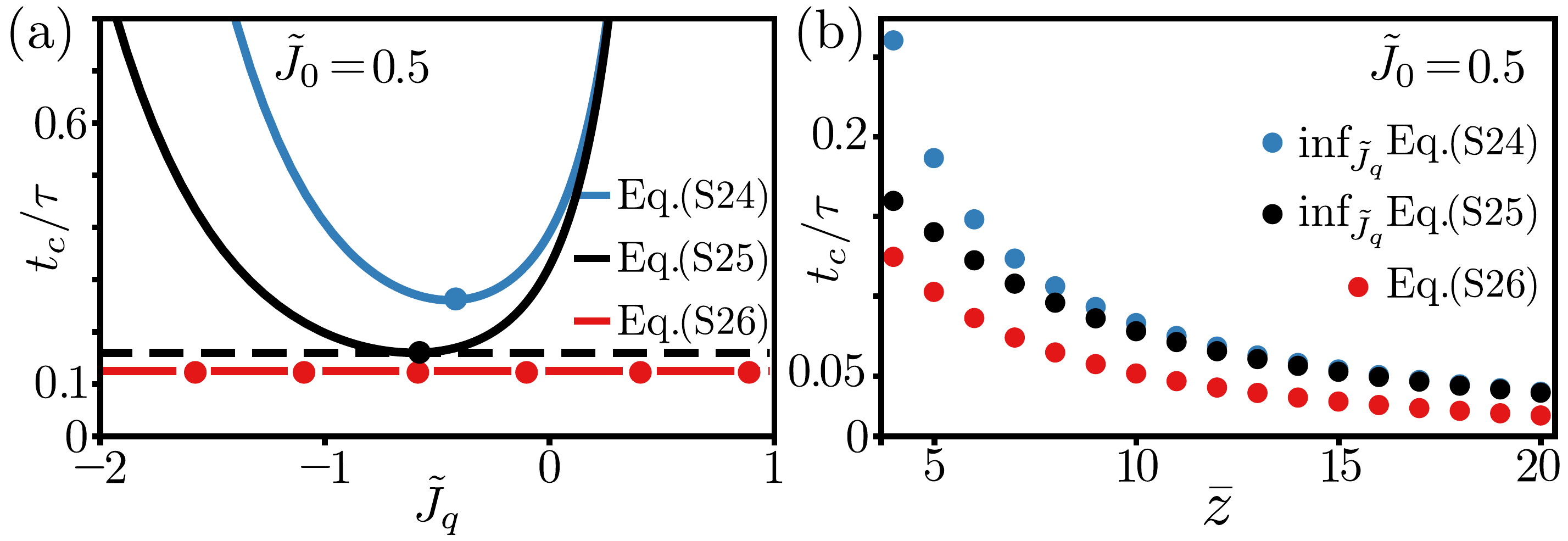}
    \caption{\textbf{Bounds on the BG critical time for quenches in
        the one-phase domain.} (a) BG critical time $t^{\rm
        BG}_{c}(\tilde{J}_{0},\tilde{J}_{q})$ given by
      Eq.~\eqref{tcBG} (blue line) as a function of the quench
      temperature $\tilde{J}_{q}$ for $\tilde{J}_{0}=0.5$ and
      $\bar{z}=4$. The respective lower bounds are shown with the
      black and red line. (b) Minimum of the BG critical time (blue
      dots) as a function of the lattice coordination number
      $\bar{z}$. The respective lower bounds are shown with the black
      and red dots, respectively.}
    \label{figSM2}
\end{figure}

\noindent For quenches in the one-phase domain we can bound the
critical time by applying the well-known inequality
$\ln{(1+x)}>2x/(2+x)$ for $x>0$ \cite{10.2307/3615890SM} to the
logarithmic term in Eq.~(7) in the Letter (since $-\tilde{\rm f}_{\rm
  BG}^{\prime\prime}(0;\tilde{J}_{q})/\tilde{\rm f}_{\rm
  BG}^{\prime\prime}(0;\tilde{J}_{0})>0$). This yields the
\emph{local} lower bound
\begin{equation}
    t^{\dagger {\rm BG}}_{c}(\tilde{J}_{0},\tilde{J}_{q})=\frac{\cosh^{\bar{z}}{(\tilde{J}_{q})}}{\bar{z}-2+\bar{z}\e{-2\tilde{J}_{q}}-2\bar{z}\e{-2\tilde{J}_{0}}}.
    \label{tcBGdagger}
\end{equation}
In Fig.~\ref{figSM2}a we plot $t^{\dagger {\rm BG}}_{c}$ with the
black line. Surprisingly, this local bound also seems to work for
$\tilde{J}_{q}\geq\tilde{J}^{\rm BG}_{c}$, even though $-\tilde{\rm f}_{\rm
  BG}^{\prime\prime}(0;\tilde{J}_{q})/\tilde{\rm f}_{\rm
  BG}^{\prime\prime}(0;\tilde{J}_{0})<0$. Furthermore, it gives the
exact result for $\tilde{J}_{q}=\tilde{J}^{\rm BG}_{c}$ given by
Eq.~\eqref{tcBGcrit}. The lower bound is also non-monotonic
w.r.t.~$\tilde{J}_{q}$, and displays a minimum for an
anti-ferromagnetic quench $\tilde{J}_{q}<0$ which we show in the next section. At the respective
minimum, the \emph{global} lower bound $\inf_{\tilde{J}_{q}}t^{\dagger {\rm BG}}_{c}(\tilde{J}_{0},\tilde{J}_{q})$ (see black dashed line in Fig.~\ref{figSM2}a) is given by Eq.~(8) in the Letter.
Taking the limit $\tilde{J}_{0}\rightarrow\infty$ of
Eq.~(8), we further obtain the following universal global lower bound
independent of $\tilde{J}_{q}$ and $\tilde{J}_{0}$ that reads
\begin{equation}
    \lim\limits_{\tilde{J}_{0}\rightarrow\infty}\inf_{\tilde{J}_{q}} t^{\dagger {\rm BG}}_{c}(\tilde{J}_{0},\tilde{J}_{q})=\frac{(\bar{z}-2)^{1-\bar{z}/2}[2+\nu_{\bar{z}}]^{-\bar{z}/2}[\bar{z}+\nu_{\bar{z}}]^{\bar{z}}}{2^{\bar{z}}(4+\bar{z}[\bar{z}-2+\nu_{\bar{z}}])},
    \label{tcBGdagger2}
\end{equation}
with $\nu_{\bar{z}}\equiv\sqrt{8+\bar{z}(\bar{z}-4)}$. For $\bar{z}=4$
this gives the universal global lower bound $t^{\dagger {\rm
    BG}}_{c}(\tilde{J}_{0},\tilde{J}_{q})>1/8$ and is shown with the
red line in Fig.~\ref{figSM2}a. In Fig.~\ref{figSM2}b we observe that for increasing $\bar{z}$ the bounds given by Eq.~(8) in the Letter and Eq.~\eqref{tcBGdagger2} become sharper with respect to the true/exact minimum of $t^{\rm BG}_{c}$. 
\subsubsection{Proof that the minimum is attained at an antiferromagnetic coupling}
It is to show that
\begin{equation}
    \tilde{J}_{q}^{\dagger}(\tilde{J}_{0})\equiv \arg\min
    _{\tilde{J}_{q}} 
    t^{\dagger}_{c}(\tilde{J}_{0},\tilde{J}_{q})=\frac{1}{2} \ln \left(
      \frac{\bar{z}-2}{\bar{z}{\rm
          e}^{-2\tilde{J}_0}\!%
        +\!
        2\!+
       \Delta_{\bar{z}}(\tilde{J}_{0})
    }
    \right)\le 0
    \label{main}
\end{equation}
with $\Delta_{\bar{z}}(\tilde{J}_{0})\equiv[8{+}\bar{z}^{2}\e{-4\tilde{J}_{0}}{+}\bar{z}(\bar{z}{-}4)(1{-}2\e{-2\tilde{J}_{0}})]^{\frac{1}{2}}$ for $\tilde{J}_0\ge\tilde{J}^{\rm BG}_{c}\equiv\ln{(\bar{z}/(\bar{z}-2))}/2$. Since 
\begin{equation}
    \lim_{\tilde{J}_0\to\infty}\tilde{J}_{q}^{\dagger}(\tilde{J}_{0})=\frac{1}{2}\ln{\left(\frac{\bar{z}-2}{2+\sqrt{\bar{z}^2-4\bar{z}+8}}\right)}\le
\frac{1}{2}\ln{\left(\frac{\bar{z}-2}{\sqrt{\bar{z}^{2}-4\bar{z}+4}}\right)}=0,
\end{equation}
and because
$\tilde{J}_{q}^{\dagger}(\tilde{J}_{0})$ is a continuously
differentiable function of $\tilde{J}_0$ for $\tilde{J}_0\ge\tilde{J}^{\rm BG}_{c}$, it suffices to show that
$d\tilde{J}_{q}^{\dagger}(\tilde{J}_{0})/d\tilde{J}_{0}\ge 0$ for $\tilde{J}_{0}\ge\tilde{J}^{\rm BG}_{c}$. 
To that aim let us differentiate Eq.~\eqref{main} w.r.t.~$\tilde{J}_{0}$:
\begin{equation}
\frac{d\tilde{J}_{q}^{\dagger}}{d\tilde{J}_{0}}=-\frac{1}{2}\frac{-2\bar{z}\e{-2\tilde{J}_{0}}+\Delta'_{\bar{z}}(\tilde{J}_{0})}{\bar{z}\e{-2\tilde{J}_{0}}+2+\Delta_{\bar{z}}(\tilde{J}_{0})}.
\label{derivative}
\end{equation}  
Noting that $\bar{z}\e{-2\tilde{J}_{0}}+2+\Delta_{\bar{z}}(\tilde{J}_{0})>0$ for all
 $\tilde{J}_0\ge\tilde{J}^{\rm BG}_{c}$, it remains to be shown that $-2\bar{z}\e{-2\tilde{J}_{0}}+\Delta'_{\bar{z}}(\tilde{J}_{0})\le 0$. Evaluating the derivative of $\Delta_{\bar{z}}(\tilde{J}_{0})$ we find
\begin{equation}
\Delta'_{\bar{z}}(\tilde{J}_{0})=\frac{2\bar{z}(\bar{z}-4)\e{-2\tilde{J}_{0}}-2\bar{z}^{2}\e{-4\tilde{J}_{0}}}{\sqrt{8{+}\bar{z}^{2}\e{-4\tilde{J}_{0}}{+}\bar{z}(\bar{z}{-}4)(1{-}2\e{-2\tilde{J}_{0}})}},
\end{equation}
which means that to prove $d\tilde{J}_{q}^{\dagger}(\tilde{J}_{0})/d\tilde{J}_{0}\ge 0$ for $\tilde{J}_{0}\ge\tilde{J}^{\rm BG}_{c}$ we must only show that
\begin{equation}
    \frac{\bar{z}-4-\bar{z}\e{-2\tilde{J}_{0}}}{\sqrt{8{+}\bar{z}^{2}\e{-4\tilde{J}_{0}}{+}\bar{z}(\bar{z}{-}4)(1{-}2\e{-2\tilde{J}_{0}})}}\leq1
    \implies (\bar{z}-4-\bar{z}\e{-2\tilde{J}_{0}})^{2}\leq8{+}\bar{z}^{2}\e{-4\tilde{J}_{0}}{+}\bar{z}(\bar{z}{-}4)(1{-}2\e{-2\tilde{J}_{0}}).
\label{only}
\end{equation}
Writing out the right hand side and canceling terms which are on both sides we finally obtain $4(2-\bar{z})\leq0$, which is indeed the case for $\bar{z}\geq2$. Hence, since we have shown that $J_{q}^\dagger(\tilde{J}_{0})$ is
monotonically increasing for $\tilde{J}_{0}\ge\tilde{J}^{\rm BG}_{c}$, and since  $\lim_{\tilde{J}_0\to\infty}\tilde{J}_{q}^{\dagger}(\tilde{J}_{0})<0$, the lower bound in
Eq.~\eqref{main} follows immediately.
\subsection{$\tilde{J}_{q}\geq\tilde{J}^{\rm BG}_{c}$}
\noindent For quenches in the two-phase domain we prove that the BG critical time $t^{\rm BG}_{c}(\tilde{J}_{0},\tilde{J}_{q})$ is bounded from below by the critical quench $t^{\rm BG}_{c}(\tilde{J}_{0},\tilde{J}^{\rm BG}_{c})$, which reads
\begin{equation}
    t^{\rm BG}_{c}(\tilde{J}_{0},\tilde{J}^{\rm BG}_{c})=\left(\frac{\bar{z}-1}{\sqrt{\bar{z}(\bar{z}-2)}}\right)^{\bar{z}}\frac{\tanh{(\tilde{J}_{0})}+1}{4((\bar{z}-1)\tanh{(\tilde{J}_{0})}-1)}.
    \label{tcBGcrit}
\end{equation}
To prove that Eq.~\eqref{tcBGcrit} provides a lower bound for the critical time for quenches in the two-phase domain, we first differentiate Eq.~\eqref{tcBG} w.r.t.~$\tilde{J}_{q}$, which gives
\begin{equation}
    \frac{\partial t^{\rm BG}_{c}(\tilde{J}_{0},\tilde{J}_{q})}{\partial \tilde{J}_{q}}=\frac{\bar{z}(1+\tanh{(\tilde{J}_{q})})\cosh^{\bar{z}}{(\tilde{J}_{q})}}{4((\bar{z}-1)\tanh{(\tilde{J}_{q})}-1)^{2}}\mathcal{A}_{1}(\tilde{J}_{0},\tilde{J}_{q}),
    \label{Eq1}
\end{equation}
where we have introduced the auxiliary function (and subsequent auxiliary functions)
\begin{eqnarray}
    \mathcal{A}_{1}(\tilde{J}_{0},\tilde{J}_{q})&\equiv&-\mathcal{A}_{2}(\tilde{J}_{0},\tilde{J}_{q})[1-\tanh{(\tilde{J}_{q})}]-\mathcal{A}_{3}(\tilde{J}_{q})\ln{(\mathcal{A}_{2}(\tilde{J}_{q},\tilde{J}_{0}))}, \nonumber \\
    \mathcal{A}_{2}(\tilde{J}_{0},\tilde{J}_{q})&\equiv&[(\bar{z}{-}1)\tanh{(\tilde{J}_{q})}{-}1][1+\tanh{(\tilde{J}_{0})}]/[\bar{z}(\tanh{(\tilde{J}_{q})}-\tanh{(\tilde{J}_{0})})],\nonumber \\
    \mathcal{A}_{3}(\tilde{J}_{q})&\equiv&1{-}(\bar{z}{-}1)\tanh^{2}{(\tilde{J}_{q})}.
    \label{Eq2}
\end{eqnarray}
All terms in front of $\mathcal{A}_{1}(\tilde{J}_{0},\tilde{J}_{q})$ in Eq.~\eqref{Eq1} are trivially positive. If furthermore $\mathcal{A}_{1}(\tilde{J}_{0},\tilde{J}_{q})>0$ for $\tilde{J}^{\rm BG}_{c}<\tilde{J}_{q}<\tilde{J}_{0}$, then we know that Eq.~\eqref{tcBGcrit} provides a lower bound. To prove that the latter is positive we proceed in two steps. 

\subsubsection{$\mathcal{A}_{2}(\tilde{J}_{q},\tilde{J}_{0})>1 \ \forall \tilde{J}_{q}>\tilde{J}^{\rm BG}_{c}$}
\noindent First we focus on the term $\mathcal{A}_{2}(\tilde{J}_{q},\tilde{J}_{0})$ entering the logarithm in $\mathcal{A}_{1}(\tilde{J}_{0},\tilde{J}_{q})$. Here we prove that $\mathcal{A}_{2}(\tilde{J}_{q},\tilde{J}_{0})>1 \ \forall \tilde{J}_{q}>\tilde{J}^{\rm BG}_{c}$, which we need for the second step. First, note that $\mathcal{A}_{2}(\tilde{J}^{\rm BG}_{c},\tilde{J}_{0})=1$, which can easily be checked by hand. Introducing $x_{0}\equiv\tanh{(\tilde{J}_{0})}$ and $x_{q}\equiv\tanh{(\tilde{J}_{q})}$, we find $\partial_{\tilde{J}_{q}}\mathcal{A}_{2}(\tilde{J}_{q},\tilde{J}_{0})=\cosh^{-2}(\tilde{J}_{q})\partial_{x_{q}}\mathcal{A}_{2}(x_{q},x_{0})>0 \ \forall \tilde{J}_{0}>\tilde{J}^{\rm BG}_{c}$. To see this, we write out the partial derivative and obtain
\begin{equation}
    \partial_{x_{q}}\mathcal{A}_{2}(x_{q},x_{0})=\partial_{x_{q}}\left(\frac{(1+x_{q})[(\bar{z}{-}1)x_{0}{-}1]}{\bar{z}(x_{0}-x_{q})}\right)=\frac{(1+x_{0})((\bar{z}-1)x_{0}-1)}{\bar{z}(x_{0}-x_{q})^{2}}>0 \ \forall x_{0}>(\bar{z}-1)^{-1}.
\end{equation}
Finally, note that $x_{0}>(\bar{z}-1)^{-1}$ translates to $\tilde{J}_{0}>\arctanh{(\bar{z}-1)^{-1}}=\tilde{J}^{\rm BG}_{c}$, which is the regime of interest. Hence, $\mathcal{A}_{2}(\tilde{J}_{q},\tilde{J}_{0})$ has a positive slope w.r.t.~$\tilde{J}_{q}$. Combined with $\mathcal{A}_{2}(\tilde{J}^{\rm BG}_{c},\tilde{J}_{0})=1$, this proves that $\mathcal{A}_{2}(\tilde{J}_{q},\tilde{J}_{0})>1 \ \forall \tilde{J}_{q}>\tilde{J}^{\rm BG}_{c}$.
\subsubsection{$\mathcal{A}_{1}(\tilde{J}_{0},\tilde{J}_{q})>0 \ \forall \tilde{J}_{q}>\tilde{J}^{\rm BG}_{c}$}
\noindent Now we turn our attention to $\mathcal{A}_{1}(\tilde{J}_{0},\tilde{J}_{q})$. We begin by considering the regime $\tanh(\tilde{J}_{q})\ge1/\sqrt{z-1}$. Here $\mathcal{A}_{3}(\tilde{J}_{q})<0$, and therefore $-\mathcal{A}_{3}(\tilde{J}_{q})\ln{(\mathcal{A}_{2}(\tilde{J}_{q},\tilde{J}_{0}))}>0$ based on the previous step. Furthermore, $-\mathcal{A}_{2}(\tilde{J}_{0},\tilde{J}_{q})[1-\tanh{(\tilde{J}_{q})}]>0 \ \forall \tilde{J}^{\rm BG}_{c}<\tilde{J}_{q}<\tilde{J}_{0}$, and so it follows that $\mathcal{A}_{1}(\tilde{J}_{0},\tilde{J}_{q})>0$ for $1/\sqrt{z-1}\leq\tanh{(\tilde{J}_{q}})\leq\tanh{(\tilde{J}_{0}})$.\vspace{0.2cm}\\
\indent Next we consider the regime $1/(z{-}1){<}\tanh(\tilde{J}_{q}){<}1/\sqrt{z{-}1}$. Here $\mathcal{A}_{3}(\tilde{J}_{q}){>}0$, and therefore ${-}\mathcal{A}_{3}(\tilde{J}_{q})\ln{(\mathcal{A}_{2}(\tilde{J}_{q},\tilde{J}_{0}))}{<}0$. To construct a bound for $\mathcal{A}_{1}(\tilde{J}_{0},\tilde{J}_{q})$ we apply the following chain of inequalities
\begin{eqnarray*}
    \mathcal{A}_{1}(\tilde{J}_{0},\tilde{J}_{q})&\equiv&\mathcal{A}_{3}(\tilde{J}_{q})[-\mathcal{A}_{2}(\tilde{J}_{0},\tilde{J}_{q})[1-\tanh{(\tilde{J}_{q})}]/\mathcal{A}_{3}(\tilde{J}_{q})-\ln{(\mathcal{A}_{2}(\tilde{J}_{q},\tilde{J}_{0}))}]\\
    &>&\mathcal{A}_{3}(\tilde{J}_{q})[-\mathcal{A}_{2}(\tilde{J}_{0},\tilde{J}_{q})[1-\tanh{(\tilde{J}_{q})}]/\mathcal{A}_{3}(\tilde{J}_{q})-\mathcal{A}_{2}(\tilde{J}_{q},\tilde{J}_{0})+1]\\
    &>&\mathcal{A}_{3}(\tilde{J}_{q})[-\mathcal{A}_{2}(\tilde{J}_{0},\tilde{J}_{q})-\mathcal{A}_{2}(\tilde{J}_{q},\tilde{J}_{0})+1]=0.
     \end{eqnarray*}
In passing from the first to the second line we have applied the
inequality $\ln{(z)}<z-1$ for $z>1$. From the second to the third line
we have used
$[1-\tanh{(\tilde{J}_{q})}]/\mathcal{A}_{3}(\tilde{J}_{q})>1$ for
$1/(z{-}1){<}\tanh(\tilde{J}_{q}){<}1/\sqrt{z{-}1}$. Finally, in the
last line we used that
$1-\mathcal{A}_{2}(x_{0},x_{q})-\mathcal{A}_{2}(x_{q},x_{0})=0$, which
follows by simply writing out the terms.\\
\indent Combining the results we find that $\mathcal{A}_{1}(\tilde{J}_{0},\tilde{J}_{q})>0$ for $\tilde{J}^{\rm BG}_{c}<\tilde{J}_{q}<\tilde{J}_{0}$, and therefore $t^{\rm BG}_{c}(\tilde{J}_{0},\tilde{J}_{q})$ is bounded by Eq.~\eqref{tcBGcrit} in this regime. 
\section{Relaxation dynamics}\label{rel}
\noindent In this section we focus on the relaxation dynamics of the
minima of the rate function, $\bar{m}(t,\tilde{J}_{0},\tilde{J}_{q})\equiv\arg\min_{m}V(m;\tilde{J}_{q},t)$, which we need to evaluate the relative entropy in Sec.~\ref{relent}. Based on the first characteristic equation in Eq.~\eqref{Hamilton} we find that the minima obey the differential equation
\begin{equation}
    \frac{d\bar{m}(t,\tilde{J}_{0},\tilde{J}_{q})}{dt}=2w^{+}(\bar{m};\tilde{J}_{q})-2w^{-}(\bar{m};\tilde{J}_{q}).
    \label{barm}
\end{equation}
As the right-hand side (RHS) does not depend explicitly on time, the
solution is given by the integral 
\begin{equation}
    \frac{1}{2}\int\frac{d\bar{m}}{w^{+}(\bar{m};\tilde{J}_{q})-w^{-}(\bar{m};\tilde{J}_{q})}=t+\mathcal{C},
    \label{implicit}
\end{equation}
where $\mathcal{C}=\mathcal{C}(\tilde{J}_{0},\tilde{J}_{q})$ is an
integration constant left to be determined from the initial condition
at $t=0$. The integral on the left-hand side (LHS) cannot be evaluated
analytically upon inserting the MF transition rates (see Eq.~(3) in
\cite{PhysRevLett.128.110603SM} for their functional form).  However,
for the BG transition rates given by Eq.~(4) in the Letter, the
integral can be evaluated explicitly for $\bar{z}=\{2,3,4,5,6\}$. Here
we show the analysis for $\bar{z}=\{2,4\}$,  where we use the former
as an educative introduction to carry out the latter. Our aim is to go
beyond the linear response regime studied in \cite{Saito1976SM} by
applying the so-called \emph{Lagrange Inversion Theorem}. 
\subsection{BG approximation with $\bar{z}=2$}
\noindent Formally the mean magnetization for $\bar{z}=2$ vanishes for
any initial and final temperature. However, instead of considering a
temperature quench, we consider a magnetization quench where we
initially prepare the system in a non-zero magnetic state with
$\bar{m}(0)\equiv\bar{m}_{0}\neq0$. Inserting the BG transition rates
with $\bar{z}=2$ into Eq.~\eqref{implicit} we obtain -- after some
algebraic manipulation -- the result
\begin{equation}
    -\tau_{r}(\tilde{J}_{q})\ln{(\bar{m}/g(\bar{m};\tilde{J}_{q}))}
    =t+\mathcal{C},
    \label{implicit1d}
\end{equation}
where $1/\tau_{r}(\tilde{J}_{q})\equiv4w^{\pm}_{\rm BG}(0;\tilde{J}_{q})\tilde{\rm f}^{\prime\prime}_{\rm BG}(0;\tilde{J}_{q})=8/(1+\e{2\tilde{J}_{q}})^{2}$ is the relaxation rate for $\bar{z}=2$, and we have introduced the auxiliary function
\begin{equation}
    g(\bar{m};\tilde{J}_{q})\equiv\exp{(-\tanh{(\tilde{J}_{q})}\ln{(\alpha_{+})}-\alpha_{-}/(2\cosh{(\tilde{J}_{q})}\bar{m})^{2})},
\end{equation}
with
\begin{equation}
    \alpha_{\pm}(\bar{m};\tilde{J}_{q})\equiv\exp{(2\tilde{J}_{q})}\pm [\bar{m}^{2}{+}\exp{(4\tilde{J}_{q})}(1{-}\bar{m}^{2})]^{1/2}.
    \label{alpha}
\end{equation}
From Eq.~\eqref{implicit1d} we directly read off the integration constant $\mathcal{C}=\mathcal{C}(\bar{m}_{0},\tilde{J}_{q})$ at $t=0$. To obtain an explicit solution for $\bar{m}$ we multiply both sides of Eq.~\eqref{implicit1d} by $-\tau_{r}$, and subsequently exponentiate, resulting in 
\begin{equation}
    \frac{\bar{m}}{g(\bar{m};\tilde{J}_{q})}=\frac{\bar{m}_{0}}{g(\bar{m}_{0};\tilde{J}_{q})}\e{-t/\tau_{r}(\tilde{J}_{q})},
    \label{implicit1d2}
\end{equation}
where we have now also fixed the integration constant. Now we invoke the \emph{Lagrange inversion theorem}: Let $f(w)$ be analytic in some neighborhood of the point $w=0$
(of the complex plane) with $f(0)\ne 0$ and let it satisfy the equation
 \begin{equation}
   \frac{w}{f(w)}=\xi.
\label{Lagrange}   
\end{equation}
Then $\exists a,b\in\mathbbm{R}^+$ such that for $|\xi| < a$ Eq.~\eqref{Lagrange} has only a single solution in the domain $|w| <
b$. According to the Lagrange-B\"urmann formula this
unique solution is an analytical function of $\xi$ given by
\begin{equation}
    w=\sum_{k=1}^{\infty}\frac{\xi^k}{k!}\left[\frac{d^{k-1}}{dw^{k-1}}f(w)^k\right]_{w=0}.
    \label{Burmann}  
\end{equation}  
Note that Eq.~\eqref{implicit1d2} is similar in structure to Eq.~\eqref{Lagrange}, and furthermore 
\begin{equation}
    g(0;\tilde{J}_{q})=\exp{(-\tanh{(\tilde{J}_{q})}(1/2{+}\ln{2}{+}2\tilde{J}_{q}))},
\end{equation} 
which is non-zero $\forall \tilde{J}_{q}\in\mathbb{R}$. Therefore, we
can use Eq.~\eqref{Burmann} to obtain an explicit solution for
$\bar{m}$, yielding
\begin{eqnarray}
    \bar{m}(t,\bar{m}_{0},\tilde{J}_{q})=\sum_{k=1}^{\infty}\frac{\bar{m}^{k}_{0}}{g(\bar{m}_{0};\tilde{J}_{q})^{k}k!}\left[\frac{d^{k-1}}{dw^{k-1}}g(\bar{m};\tilde{J}_{q})^{k}\right]_{\bar{m}=0}\e{-kt/\tau_{r}(\tilde{J}_{q})}
     =\sum_{k=1}^{\infty}\alpha_{k}(\bar{m}_{0};\tilde{J}_{q})\e{-kt/\tau_{r}(\tilde{J}_{q})}.
    \label{sol}
\end{eqnarray}
For completeness, we list the first three non-zero coefficients
\begin{eqnarray}
    \alpha_{1}(\bar{m}_{0};\tilde{J}_{q})&=&\bar{m}_{0}g(0;\tilde{J}_{q})/g(\bar{m}_{0};\tilde{J}_{q}),\nonumber\\
    \alpha_{3}(\bar{m}_{0};\tilde{J}_{q})&=&\alpha^{3}_{1}(\bar{m}_{0};\tilde{J}_{q})\e{-4\tilde{J}_{q}}(1-\e{2\tilde{J}_{q}})^{2}/8,\nonumber\\
    \alpha_{5}(\bar{m}_{0};\tilde{J}_{q})&=&\alpha^{5}_{1}(\bar{m}_{0};\tilde{J}_{q})\e{-4\tilde{J}_{q}}\sinh{(\tilde{J}_{q})}^{3}(4\cosh{(\tilde{J}_{q})}+5\sinh{(\tilde{J}_{q})})/8.
\end{eqnarray}
Note that $\alpha_{1}(\bar{m}_{0};0)=\bar{m}_{0}$ and
$\alpha_{k}(\bar{m}_{0};0)=0 \ \forall k\in\{2,3,...\}$, which gives
the well-known result
$\bar{m}(t,\bar{m}_{0},0)=\bar{m}_{0}\exp{(-2t)}$
\cite{singu_2SM}. Furthermore, since
$g(\bar{m};\tilde{J}_{q})=g(-\bar{m};\tilde{J}_{q})$, we know that
$\alpha_{2k}=0 \ \forall k \in \mathbb{N}$. This concludes our
derivation of $\bar{m}(t,\bar{m}_{0},\tilde{J}_{q})$ for $\bar{z}=2$. 
\subsection{BG approximation with $\bar{z}=4$}
\noindent Now we focus on the case $\bar{z}=4$. The analysis requires the same steps as shown in the previous section, but involves a bit more algebra. We will focus only on quenches where the initial temperature is below the critical temperature, i.e.~$\tilde{J}_{0}>\tilde{J}^{\rm BG}_{c}=\ln{(2)}/2$, resulting in the following initial magnetization \cite{blom2021criticalitySM}
\begin{equation}
    \bar{m}_{0}(\tilde{J}_{0})=\e{2\tilde{J}_{0}}(\e{4\tilde{J}_{0}}-4)^{1/2}/(\e{4\tilde{J}_{0}}-2).
    \label{m0}
\end{equation}
 In order to apply the \emph{Lagrange inversion theorem} we have to
 make a distinction between quenches above and below the critical
 temperature, since they have different equilibrium states.
 Furthermore, for quenches above the critical temperature
 $\tilde{J}_{q}\leq\ln{(2)}/2$, we will encounter a particular
 ``special'' value
 $\tilde{J}_{q}=\ln{(2)}/4$ which needs to be handled separately.  
\subsubsection{$\tilde{J}_{q}<\ln{(2)}/2$ and $\tilde{J}_{q}\neq\ln{(2)}/4$}
\noindent Upon determining the integral in Eq.~\eqref{implicit} for
$\bar{z}=4$ we obtain an analytic expression which can be written in a
similar form as Eq.~\eqref{implicit1d}. In this regime the relaxation
rate is given by $1/\tau_{r}(\tilde{J}_{q})\equiv4w^{\pm}_{\rm BG}(0;\tilde{J}_{q})\tilde{\rm f}^{\prime\prime}_{\rm
  BG}(0;\tilde{J}_{q})=\cosh^{4}{(\tilde{J}_{q})}/(4\exp{(-2\tilde{J})}-2)$,
which is plotted in Fig.~2d in the Letter  (green + blue line). The auxiliary function
$g(\bar{m};\tilde{J}_{q})$ in Eq.~\eqref{implicit1d} is now given by 
\begin{equation}
    g(\bar{m};\tilde{J}_{q})=\prod_{i=1}^{5}g_{i}(\bar{m};\tilde{J}_{q}),
    \label{g2d1}
\end{equation}
which we have further divided into sub-auxiliary functions that read
\begin{eqnarray}
    g_{1}(\bar{m};\tilde{J}_{q})&=&\exp{\left(\frac{\alpha_{-}\sech{(\tilde{J}_{q})}^{6}(1-3\tanh{(\tilde{J}_{q})})(2+\e{2\tilde{J}_{q}})^{2}}{8m^{4}(\tanh{(\tilde{J}_{q})}-3)^{3}}\right)},\nonumber\\
    g_{2}(\bar{m};\tilde{J}_{q})&=&\exp{\left(\frac{\e{2\tilde{J}_{q}}(2-\e{2\tilde{J}_{q}})(2\alpha_{-}+(13\alpha_{-}-2)\e{2\tilde{J}_{q}}+(5\alpha_{-}+1)\e{4\tilde{J}_{q}}+\e{6\tilde{J}_{q}})}{(1+\e{2\tilde{J}_{q}})^{3}(2+\e{2\tilde{J}_{q}})^{2}m^{2}}\right)},\nonumber\\
    g_{3}(\bar{m};\tilde{J}_{q})&=&\alpha_{+}(\bar{m};\tilde{J}_{q})^{\nu_{1}(\tilde{J}_{q})},\nonumber\\
    \nonumber\\
    g_{4}(\bar{m};\tilde{J}_{q})&=&[4\bar{m}^{2}-\e{4\tilde{J}_{q}}(\e{4\tilde{J}_{q}}-4)(1-\bar{m}^{2})]^{\nu_{2}(\tilde{J}_{q})},\nonumber\\
    \nonumber\\
    g_{5}(\bar{m};\tilde{J}_{q})&=&[4\bar{m}^{2}+\e{4\tilde{J}_{q}}((2-\alpha_{+}+\e{2\tilde{J}_{q}}-\e{4\tilde{J}_{q}})^{2}-\bar{m}^{2}(3-\e{4\tilde{J}_{q}})^{2})]^{-\nu_{2}(\tilde{J}_{q})/2},
\end{eqnarray}
and $\alpha_{\pm}(\bar{m};\tilde{J}_{q})$ is given by
Eq.~\eqref{alpha}. The exponents in the last three equations are given
by 
\begin{eqnarray}
    \nu_{1}(\tilde{J}_{q})&\equiv&[44\tanh{(\tilde{J}_{q})}-20+\sech{(\tilde{J}_{q})}^{4}(3\tanh{(\tilde{J}_{q})}-1)+\sech{(\tilde{J}_{q})}^{2}(19\tanh{(\tilde{J}_{q})}-11)](\tanh{(\tilde{J}_{q})}-3)^{-3},\nonumber\\
    \nu_{2}(\tilde{J}_{q})&\equiv&32\e{2\tilde{J}_{q}}(2+\e{2\tilde{J}_{q}})^{-3}(\e{4\tilde{J}_{q}}-2)^{-1}.
\end{eqnarray}
Note that $\nu_{2}\rightarrow\infty$
for $\tilde{J}_{q}\rightarrow\ln{(2)}/4<\ln{(2)}/2$, which
is a particular point where the integral Eq.~\eqref{implicit}
drastically simplifies as we will see in the next section.  To check whether we can apply the \emph{Lagrange inversion theorem} we first need to determine $g(0;\tilde{J}_{q})$, which results in
\begin{equation}
    g(0;\tilde{J}_{q})=2^{\nu_{1}(\tilde{J}_{q})-\nu_{2}(\tilde{J}_{q})}\exp{(2[\nu_{1}(\tilde{J}_{q})+\nu_{2}(\tilde{J}_{q})]\tilde{J}_{q}-\nu_{3}(\tilde{J}_{q}))}|\coth{(\tilde{J}_{q})}-3|^{\nu_{2}(\tilde{J}_{q})},
\end{equation}
where we have defined the auxiliary function
\begin{equation}
    \nu_{3}(\tilde{J}_{q})=(9\e{8\tilde{J}_{q}}-2\e{6\tilde{J}_{q}}-51\e{4\tilde{J}_{q}}+32\e{2\tilde{J}_{q}}+12)/4(\e {4\tilde{J}_{q}}+3\e{2\tilde{J}_{}q}+2)^{2}.
\end{equation}
For $\tilde{J}_{q}<\ln{(2)}/2$ \emph{and}
$\tilde{J}_{q}\neq\ln{(2)}/4$ we have $\coth{(\tilde{J}_{q})}-3\neq0$
and $|\nu_{1,2,3}(\tilde{J}_{q})|<\infty$. Hence, in this regime
$g(0;\tilde{J}_{q})\neq0$, and therefore we can use the \emph{Lagrange
inversion theorem} as in the previous section. Plugging
$g(\bar{m};\tilde{J}_{q})$ given by Eq.~\eqref{g2d1} into
Eq.~\eqref{sol}, and using Eq.~\eqref{m0} to express $\bar{m}_{0}$ in
terms of $\tilde{J}_{0}$, we obtain a power series solution. For completeness, we list the first
three non-zero coefficients 
\begin{eqnarray}
    \alpha_{1}(\tilde{J}_{0},\tilde{J}_{q})&=&\bar{m}_{0}g(0;\tilde{J}_{q})/g(\bar{m}_{0};\tilde{J}_{q}),\nonumber\\
    \alpha_{3}(\tilde{J}_{0},\tilde{J}_{q})&=&\alpha^{3}_{1}(\tilde{J}_{0},\tilde{J}_{q})\e{-4\tilde{J}_{q}}(4-\e{2\tilde{J}_{q}})(1-\e{2\tilde{J}_{q}})^{2}/(4(2-\e{2\tilde{J}_{q}})),\nonumber\\
    \alpha_{5}(\tilde{J}_{0},\tilde{J}_{q})&=&\alpha^{5}_{1}(\tilde{J}_{0},\tilde{J}_{q})\frac{111\cosh{(\tilde{J}_{q})}{-}87\cosh{(3\tilde{J}_{q})}{-}313\sinh{(\tilde{J}_{q})}{+}113\sinh{(3\tilde{J}_{q})}}{8(\coth{(\tilde{J}_{q})}-3)^{2}}\e{-4\tilde{J}_{q}}\sinh{(\tilde{J}_{q})}.
    \label{alpha1}
\end{eqnarray}
Note that only terms of $\bar{m}^{2}$ and $\bar{m}^{4}$ enter in
$g(\bar{m};\tilde{J}_{q})$ given by Eq.~\eqref{g2d1}. Therefore
$g(\bar{m};\tilde{J}_{q})=g(-\bar{m};\tilde{J}_{q})$, which implies
that $\alpha_{2k}=0 \ \forall k \in \mathbb{N}$. Furthermore, we also
have $\alpha_{1}(\tilde{J}_{0},0)=1$ and
$\alpha_{k}(\tilde{J}_{0},0)=0 \ \forall k\in\{2,3,...\}$ as in the
previous section.
\subsubsection{$\tilde{J}_{q}=\ln{(2)}/4$}
\noindent For $\tilde{J}_{q}=\ln{(2)}/4$ the outcome of the integral in Eq.~\eqref{implicit1d} simplifies drastically, and the resulting expression for the auxiliary function $g(\bar{m};\ln{(2)}/4)$ reads
\begin{equation}
    g(m;\ln{(2)}/4)=\exp{\left(\frac{c_{1}+c_{2}m^{2}-(c_{1}+(c_{1}/4+c_{2})\bar{m}^{2}-\sqrt{2}c_{3}\bar{m}^{4})[1-\bar{m}^{2}/2]^{1/2}}{m^{4}}\right)}(2+[4-2m^{2}]^{1/2})^{c_{4}},
    \label{g2d2}
\end{equation}
with the numerical coefficients given by 
\begin{equation}
    c_{1}=560\sqrt{2}-792, \ c_{2}=1092-772\sqrt{2}, \ c_{3}=8(7-5\sqrt{2}), \ c_{4}=329-232\sqrt{2}.
\end{equation}
This function attains the following value at $\bar{m}=0$ 
\begin{equation}
    g(0;\ln{(2)}/4)=4^{c_{4}}\exp{\left(\frac{3c_{1}}{32}+\frac{c2}{4}+\sqrt{2}c_{3}\right)}.
\end{equation}
Hence, $g(0;\ln{(2)}/4)\neq0$, and therefore we can use the
\emph{Lagrange inversion theorem}. Inserting Eq.~\eqref{g2d2} into
Eq.~\eqref{sol} we obtain an expression for the coefficients. The
result for the first three non-zero coefficients reads 
\begin{eqnarray}
    \alpha_{1}(\tilde{J}_{0},\ln{(2)}/4)&=&\bar{m}_{0}g(0;\ln{(2)}/4)/g(\bar{m}_{0};\ln{(2)}/4),\nonumber\\
    \alpha_{3}(\tilde{J}_{0},\ln{(2)}/4)&=&\alpha^{3}_{1}(\tilde{J}_{0},\ln{(2)}/4)(c_{1}+2c_{2}-8(2\sqrt{2}c_{3}+c_{4}))/4^{3},\nonumber\\
    \alpha_{5}(\tilde{J}_{0},\ln{(2)}/4)&=&\alpha^{5}_{1}(\tilde{J}_{0},\ln{(2)}/4)\mathcal{A}(c_{1},c_{2},c_{3},c_{4})/2^{13},
    \label{alpha2}
\end{eqnarray}
with $\mathcal{A}(c_{1}{,}c_{2}{,}c_{3}{,}c_{4}){=}5c_{1}^{2}{+}20c_{2}^{2}{+}4c_{1}(9{+}5c_{2}{-}40\sqrt{2}c_{3}{-}20c_{4}){+}32c_{2}(2{-}10\sqrt{2}c_{3}{-}5c_{4}){+}64(40c_{3}^{2}{+}c_{4}(5c_{4}{-}3){+}4\sqrt{2}c_{3}(5c_{4}{-}1))$.
Also here we find that only terms of $\bar{m}^{2}$ and $\bar{m}^{4}$
enter in Eq.~\eqref{g2d2}, which  implies that $\alpha_{2k}=0
\ \forall k \in \mathbb{N}$. Notably, the coefficients in
Eq.~\eqref{alpha1} approach Eq.~\eqref{alpha2} in the neighborhood of
$\tilde{J}_{q}=\ln{(2)}/4$. 
\subsubsection{$\tilde{J}_{q}>\ln{(2)}/2$}
\noindent Finally, we focus on a quench in the two-phase domain with
$\tilde{J}_{q}>\ln{(2)}/2$. Formally the integral given by
Eq.~\eqref{implicit} does not change w.r.t.~the analysis for
$\tilde{J}_{q}<\ln{(2)}/2$. However, there is a difference in applying
the \emph{Lagrange inversion theorem}, since the steady-state
magnetization
$\bar{m}_{\infty}(\tilde{J}_{q})=\pm\e{2\tilde{J}_{q}}(\e{4\tilde{J}_{q}}-4)^{1/2}/(\e{4\tilde{J}_{q}-2})$
maintains a non-zero value for $\tilde{J}_{q}>\ln{(2)}/2$. The
relaxation rate now reads $1/\tau_{r}(\tilde{J}_{q})=4w^{\pm}_{\rm
  BG}(m_{\infty};\tilde{J}_{q})\tilde{\rm f}_{\rm
  BG}^{\prime\prime}(\bar{m}_{\infty};\tilde{J}_{q}){=}(\e{4\tilde{J}_{q}}-2)(\e{2\tilde{J}_{q}}-2)(\e{2\tilde{J}_{q}}+2)^{3}/(\e{4\tilde{J}_{q}}+1)^{4}$. After
some algebraic manipulation, we obtain 
\begin{equation}
    -\tau_{r}(\tilde{J}_{q})\ln{\left(\frac{\bar{m}-\bar{m}_{\infty}}{g(\bar{m};\tilde{J}_{q})}\right)}
    =t+\mathcal{C},
    \label{implicit2d}
\end{equation}
where $\mathcal{C}=\mathcal{C}(\tilde{J}_{0},\tilde{J}_{q})$ is the integration constant determined by the initial condition. The function $g(\bar{m};\tilde{J}_{q})$ reads
\begin{equation}
    g(\bar{m};\tilde{J}_{q})=(\e{4\tilde{J}_{q}}-2)^{-1}\prod_{i=1}^{5}g_{i}(\bar{m};\tilde{J}_{q}),
    \label{g2d3}
\end{equation}
which we have further divided into the following sub-auxiliary functions
\begin{eqnarray}
    g_{1}(\bar{m};\tilde{J}_{q})&=&\exp{\left(\frac{\alpha_{-}(\e{4\tilde{J}_{q}}-2)(\e{2\tilde{J}_{q}}+2)^{2}(2-\e{2\tilde{J}_{q}})\e{4\tilde{J}_{q}}}{16m^{4}(1+\e{2\tilde{J}})^{4}}\right)},\nonumber\\
    g_{2}(\bar{m};\tilde{J}_{q})&=&\exp{\left(\frac{\e{4\tilde{J}_{q}}(8-6\e{4\tilde{J}_{q}}+\e{8\tilde{J}_{q}})(14+20\e{2\tilde{J}_{q}}+6\e{4\tilde{J}_{q}}-\e{-4\tilde{J}_{q}}(\e{2\tilde{J}_{q}}+1)(2+13\e{2\tilde{J}_{q}}+5\e{4\tilde{J}_{q}})(\alpha_{+}-\e{2\tilde{J}_{q}}))}{32(1+\e{2\tilde{J}_{q}})^{4}m^{2}}\right)},\nonumber\\
    g_{3}(\bar{m};\tilde{J}_{q})&=&[\bar{m}^{2}(\e{4\tilde{J}_{q}}-2)^{2}(1-\e{4\tilde{J}_{q}})+4\e{4\tilde{J}_{q}}(1-\alpha_{+}+\e{2\tilde{J}_{q}})-\e{8\tilde{J}_{}q}(3-2\alpha_{+}+2\e{2\tilde{J}_{q}})+\e{12\tilde{J}_{q}}]^{1/2},\nonumber\\
    g_{4}(\bar{m};\tilde{J}_{q})&=&|\bar{m}(\e{4\tilde{J}_{q}}-2)+\e{2\tilde{J}_{q}}(\e{4\tilde{J}_{q}}-4)^{1/2}|^{-1},\nonumber\\
    g_{5}(\bar{m},\tilde{J}_{q})&=&\bar{m}^{\nu_{1}(\tilde{J}_{q})},\nonumber\\
    g_{6}(\bar{m};\tilde{J}_{q})&=&\alpha_{+}^{-\nu_{2}(\tilde{J}_{q})}.
\end{eqnarray}
The function $\alpha_{\pm}(\bar{m};\tilde{J}_{q})$ is given by Eq.~\eqref{alpha}, and the exponents in the last two equations are given by
\begin{align}
    &\nu_{1}(\tilde{J}_{q}){\equiv}\e{-2\tilde{J}_{q}}(\e{2\tilde{J}_{q}}+2)^{3}(\e{4\tilde{J}_{q}}-2)/32, \nonumber \\
    &\nu_{2}(\tilde{J}_{q}){\equiv}\e{\tilde{J}_{q}}(\e{4\tilde{J}_{q}}{-}2)\sech{(\tilde{J}_{q})}(28{+}31\cosh{(2\tilde{J}_{q})}{+}5\cosh{(4\tilde{J}_{q})}{-}41\sinh{(2\tilde{J}_{q})}{-}11\sinh{(4\tilde{J}_{q})}{-}6\tanh{(\tilde{J}_{q})})/64.
\end{align}
In order to apply the \emph{Lagrange inversion theorem} we need to evaluate $g(\bar{m};\tilde{J}_{q})$ at the steady state $\bar{m}_{\infty}$, which yields
\begin{equation}
    g(\bar{m}_{\infty};\tilde{J}_{q})=(\bar{m}_{\infty})^{\nu_{1}(\tilde{J}_{q})}(1+\e{2\tilde{J}_{q}}+2(\e{4\tilde{J}_{q}}-2)^{-1})^{-\nu_{2}(\tilde{J}_{q})}\e{2\tilde{J}_{q}-\nu_{3}(\tilde{J}_{q})}(\e{4\tilde{J}_{q}}-2)^{-1/2},
\end{equation}
where we have defined the auxiliary function
\begin{equation}
    \nu_{3}(\tilde{J}_{q})=\e{3\tilde{J}_{q}}(13+8\cosh{(2\tilde{J}_{q})})(\cosh{(\tilde{J}_{q})}-3\sinh{(\tilde{J}_{q})})(\cosh{(\tilde{J}_{q})}-\sinh{(\tilde{J}_{q})}(6-\tanh{(\tilde{J}_{q})}))^{2}.
\end{equation}
For $\tilde{J}_{q}>\ln{(2)}/2$ we find that $g(\bar{m}_{\infty};\tilde{J}_{q})\neq0$, and therefore we can apply the \emph{Lagrange inversion theorem}. Upon inverting Eq.~\eqref{implicit2d}, the final result reads
\begin{eqnarray}
    \bar{m}(t,\bar{m}_{0},\tilde{J}_{q})&=&\bar{m}_{\infty}+\sum_{k=1}^{\infty}\frac{(\bar{m}_{0}-\bar{m}_{\infty})^{k}}{g(\bar{m}_{0};\tilde{J}_{q})^{k}k!}\left[\frac{d^{k-1}}{dw^{k-1}}g(\bar{m};\tilde{J}_{q})^{k}\right]_{\bar{m}=\bar{m}_{\infty}}\e{-kt/\tau_{r}(\tilde{J}_{q})}\nonumber\\
    &=&\bar{m}_{\infty}+\sum_{k=1}^{\infty}\alpha_{k}(\tilde{J}_{0},\tilde{J}_{q})\e{-kt/\tau_{r}(\tilde{J}_{q})}.
    \label{sol2}
\end{eqnarray}
For completeness, we list the first three non-zero coefficients
\begin{eqnarray}
    \alpha_{1}(\tilde{J}_{0},\tilde{J}_{q})&=&(\bar{m}_{0}-\bar{m}_{\infty})g(\bar{m}_{\infty};\tilde{J}_{q})/g(\bar{m}_{0};\tilde{J}_{q}),\nonumber\\
    \alpha_{2}(\tilde{J}_{0},\tilde{J}_{q})&=&\alpha^{2}_{1}(\tilde{J}_{0},\tilde{J}_{q})\e{-6\tilde{J}_{q}}(\e{4\tilde{J}_{q}}-2)^{2}(4\e{3\tilde{J}_{q}}\sinh{(\tilde{J}_{q})}-1)/(\e{4\tilde{J}_{q}}-4)^{1/2},\nonumber\\
    \alpha_{3}(\tilde{J}_{0},\tilde{J}_{q})&=&\alpha^{3}_{1}(\tilde{J}_{0},\tilde{J}_{q})\e{-6\tilde{J}_{q}}(\e{4\tilde{J}_{q}}{-}2)^{3}(52{-}10\e{-6\tilde{J}_{q}}{-}24\e{-4\tilde{J}_{q}}{+}25\e{-2\tilde{J}_{q}}{-}35\e{2\tilde{J}_{q}}{-}18\e{4\tilde{J}_{q}}{+}11\e{6\tilde{J}_{q}})/2(\e{4\tilde{J}_{q}}{-}4).
    \label{alpha3}
\end{eqnarray}
This concludes our derivation for the relaxation dynamics of the rate function minima. 
\section{Relative entropy}\label{relent}
\noindent Here we derive the coefficients $\gamma_{k}$ for the power series expansion of the relative entropy per spin, given by Eq.~(9) in the Letter. The relative entropy is evaluated with the saddle point approximation in the thermodynamic limit, which results in 
\begin{equation}
    \mathscrsfs{D}_t{=}\lim_{N\rightarrow\infty}{\int_{-1}^{1}}\!\!\!\e{-NV(m;\tilde{J}_{q},t)}[V_{\rm eq}(m;\tilde{J}_{q}){-}V(m;\tilde{J}_{q},t)]dm\simeq V_{\rm eq}(\bar{m}(t,\tilde{J}_{0},\tilde{J}_{q}),\tilde{J}_{q})=\sum_{k=2}^{\infty}\gamma_{k}(\tilde{J}_{0},\tilde{J}_{q})\e{-kt/\tau_{r}(\tilde{J}_{q})}.
    \label{powserSM}
\end{equation}
To arrive at the second equality we have applied the saddle point
approximation around the minimum
$\bar{m}(t,\tilde{J}_{0},\tilde{J}_{q})$ of the rate function
$V(m;\tilde{J}_{q},t)$ at time $t$. Note that
$V(\bar{m};\tilde{J}_{q},t)=0$, and therefore only the equilibrium
potential $V_{\rm eq}(\bar{m};\tilde{J}_{q})$ remains after the saddle
point approximation.  For the final equality we carried out a Taylor expansion around the steady state $\bar{m}_{\infty}$, and used the power series expansion of $\bar{m}(t,\tilde{J}_{0},\tilde{J}_{q})$ which is analyzed in Sec.~\ref{rel}. 
The first three non-zero coefficients in Eq.~\eqref{powserSM} are given by
\begin{eqnarray}
    \gamma_{2}(\tilde{J}_{0},\tilde{J}_{q})&=&\alpha^{2}_{1}V_{\rm eq}^{\prime\prime}(\bar{m}_{\infty};\tilde{J}_{q})/2,\nonumber\\
    \gamma_{3}(\tilde{J}_{0},\tilde{J}_{q})&=&\alpha_{1}\alpha_{2}V_{\rm eq}^{\prime\prime}(\bar{m}_{\infty};\tilde{J}_{q})+\alpha^{3}_{1}V_{\rm eq}^{\prime\prime\prime}(\bar{m}_{\infty};\tilde{J}_{q})/6,\nonumber\\
    \gamma_{4}(\tilde{J}_{0},\tilde{J}_{q})&=&(\alpha^{2}_{2}/2+\alpha_{1}\alpha_{3})V_{\rm eq}^{\prime\prime}(\bar{m}_{\infty},\tilde{J}_{q})+\alpha^{2}_{1}\alpha_{2}V_{\rm eq}^{\prime\prime\prime}(\bar{m}_{\infty};\tilde{J}_{q})/2+\alpha^{4}_{1}V_{\rm eq}^{\prime\prime\prime\prime}(\bar{m}_{\infty};\tilde{J}_{q})/24,
\end{eqnarray}
where the coefficients
$\alpha_{i}=\alpha_{i}(\tilde{J}_{0},\tilde{J}_{q})$ are given by
Eq.~\eqref{alpha1} and \eqref{alpha3} for quenches in the one- and
two-phase domain. For quenches in the one-phase domain we have
$\gamma_{3}(\tilde{J}_{0},\tilde{J}_{q})=0$ since $\bar{m}_{\infty}=0$
and $\alpha_{2}=V_{\rm
  eq}^{\prime\prime\prime}(0;\tilde{J}_{q})=0$. The inset of Fig.~2c
in the Letter displays the first two non-zero coefficients for
quenches in the one- and two-phase domain. 
\begin{figure}[t!]
    \centering
    \includegraphics[width=\textwidth]{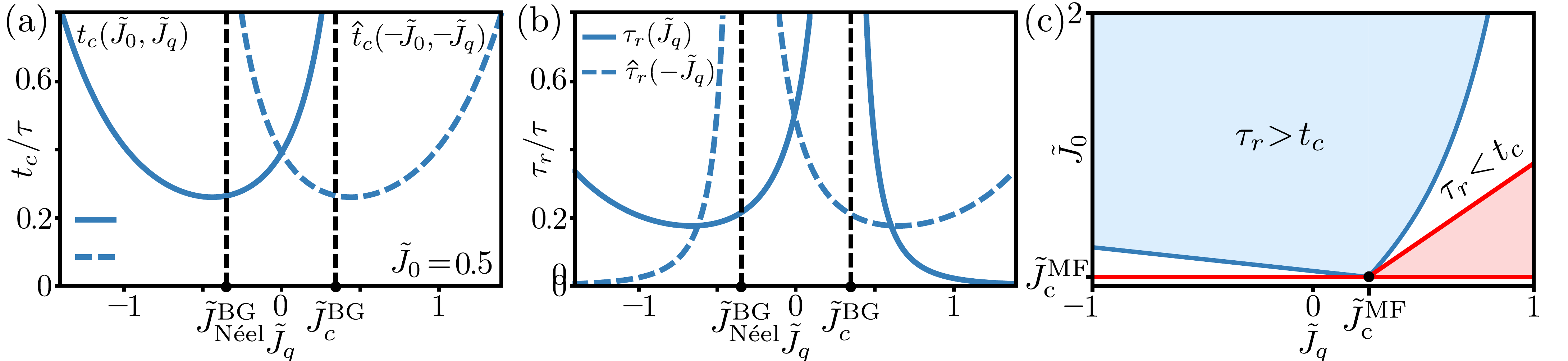}
    \caption{\textbf{Parity symmetry for the staggered magnetization and the MF dynamical phase diagram.} In all panels we consider a lattice with $\bar{z}=4$. (a)-(b) Critical time (a) and relaxation time (b) as a function of the quench temperature $\tilde{J}_{q}$. The dashed lines correspond to the staggered magnetization dynamics, for which a parity symmetry applies w.r.t. the temperature $\tilde{J}\rightarrow -\tilde{J}$ (see Eq.~\eqref{par}). (c) Dynamical phase diagram for the MF critical time
      $t^{\rm MF}_{c}$ and relaxation $\tau^{\rm MF}_{r}$ time. The red area is forbidden since
      $\tilde{J}_{0}>\tilde{J}_{q}$. Inside the blue area, the relaxation time is larger than the critical time. The dark blue phase boundary where $t^{\rm MF}_{c}=\tau^{\rm MF}_{r}$ is given by Eq.~\eqref{Jdagger}. The MF critical point reads $\tilde{J}^{\rm MF}_{c}\equiv1/\bar{z}$. Fig.~2f in the Letter shows the BG dynamical phase diagram.}
    \label{figSM3}
\end{figure}
\section{Parity symmetry for the staggered magnetization}
\noindent Let us define the staggered magnetization $\hat{m}\in[-1,1]$ in the Ising model as 
\begin{equation}
    \hat{m}\equiv N^{-1}\sum_{i=1}^{N}(-\sigma_{i})^{i}.
\end{equation}
For perfectly anti-ferromagnetic order we have $\hat{m}=\pm1$, and for anti-ferromagnetic disorder $\hat{m}=0$. 
Based on the works in \cite{10.1143/PTP.51.82SM, Ono_1984SM, Peruggi_1983SM} we know that the BG free energy density $\tilde{\rm f}_{\rm BG}(m;\tilde{J})$ obeys the following parity symmetry w.r.t.~the staggered magnetization
\begin{equation}
    \tilde{\rm f}_{\rm BG}(m;\tilde{J})=\tilde{\rm f}_{\rm BG}(\hat{m};-\tilde{J}).
    \label{par}
\end{equation}
Therefore, our results for the critical time, relaxation time, and dynamical phase diagram also apply for dynamics of staggered magnetization upon inverting the temperature $\tilde{J}\rightarrow-\tilde{J}$. In Fig.~\ref{figSM3}a-b we depict the critical time $\hat{t}_{c}$ (a) and relaxation time $\hat{\tau}_{r}$ (b) for the dynamics of the staggered magnetization with the blue dashed lines. 
\section{MF dynamical phase diagram}
\noindent Fig.~\ref{figSM3}c depicts the MF dynamical phase diagram. To obtain the blue shaded area where $\tau^{\rm MF}_{r}>t^{\rm MF}_{c}$, we first compute the MF critical time. Inserting the MF transition rates and free energy density into Eq.~(7) in the Letter we obtain the MF critical time
\begin{equation}
    t^{\rm MF}_{c}(\tilde{J}_{0},\tilde{J}_{q})=\frac{1}{4(1-\bar{z}J_{q})}\ln{\left(\frac{\bar{z}\tilde{J}_{q}-\bar{z}\tilde{J}_{0}}{1-\bar{z}\tilde{J}_{0}}\right)},
    \label{tcMF}
\end{equation}
which is also reported in \cite{PhysRevLett.128.110603SM,singu_1SM,singu_2SM} for $\bar{z}=1$. The MF relaxation time is given by $\tau^{\rm MF}_{r}(\tilde{J}_{q}){\equiv}1/4w^{\pm}_{\rm MF}(\bar{m};\tilde{J}_{q})\tilde{\rm f}^{\prime\prime}_{\rm MF}(\bar{m};\tilde{J}_{q})$, where $\bar{m}=\arg\min_{m}\tilde{\rm f}_{\rm MF}(m;\tilde{J}_{q})$ is given by the transcendental equation
\begin{equation}
    \bar{m}=\tanh{(\bar{z}\tilde{J}_{q}\bar{m})}.
\end{equation}
Equating $t^{\rm MF}_{c}$ and $\tau^{\rm MF}_{r}$ we obtain the dark blue boundary line 
\begin{equation}
    \bar{z}\tilde{J}^{\dagger}_{0}=\frac{\bar{z}\tilde{J}_{q}\exp{\left(\frac{2(1+\bar{m})(\bar{z}\tilde{J}_{q}-1)}{1-(1-\bar{m}^{2})\bar{z}\tilde{J}_{q}}\e{-\bar{z}\bar{m}\tilde{J}_{q}}\right)}-1}{\exp{\left(\frac{2(1+\bar{m})(\bar{z}\tilde{J}_{q}-1)}{1-(1-\bar{m}^{2})\bar{z}\tilde{J}_{q}}\e{-\bar{z}\bar{m}\tilde{J}_{q}}\right)}-1}.
    \label{Jdagger}
\end{equation}
For $\tilde{J}_{0}>\tilde{J}^{\dagger}_{0}$ (blue region) the MF
relaxation time is larger than the critical time, i.e. $\tau^{\rm
  MF}_{r}>t^{\rm MF}_{c}$. For
$1/\bar{z}<\tilde{J}_{0}<\tilde{J}^{\dagger}_{0}$ (white region) the
MF critical time is larger than the relaxation time. 

\end{document}